# Understanding overfitting in random forest for probability estimation: a visualization and simulation study

**Lasai Barreñada**[1], **Paula Dhiman**[2], **Dirk Timmerman**[1,3], **Anne-Laure Boulesteix**[4], **Ben Van Calster**[*,1,5,6]

1 Department of Development and Regeneration, KU Leuven, Belgium
2 Centre for Statistics in Medicine, Nuffield Department of Orthopaedics, Rheumatology and Musculoskeletal Sciences, University of Oxford, United Kingdom
3 Department of Obstetrics and Gynecology, University Hospitals Leuven, Belgium
4 Biometry in Molecular Medicine, LMU Munich, Germany
5 Leuven Unit for Health Technology Assessment Research (LUHTAR), KU Leuven, Belgium
6 Department of Biomedical Data Sciences, Leiden University Medical Centre, Leiden, the Netherlands

**ABSTRACT**:

**Background**: Random forests have become popular for clinical risk prediction modelling. In a case study on predicting ovarian malignancy, we observed training AUCs close to 1. Although this suggests overfitting, performance was competitive on test data. We aimed to understand the behavior of random forests for probability estimation by (1) visualizing data space in three real world case studies and (2) a simulation study.

**Methods**: For the case studies, multinomial risk estimates were visualized using heatmaps in a 2-dimensional subspace. The simulation study included 48 logistic data generating mechanisms (DGM), varying the predictor distribution, the number of predictors, the correlation between predictors, the true AUC and the strength of true predictors. For each DGM, 1000 training datasets of size 200 or 4000 with binary outcome were simulated and random forest models trained with minimum node size 2 or 20 using the ranger R package, resulting in 192 scenarios in total. Model performance was evaluated on large test datasets (N=100,000).

**Results**: The visualizations suggested that the model learned 'spikes of probability' around events in the training set. A cluster of events created a bigger peak or plateau (signal), isolated events local peaks (noise). In the simulation study, median training AUCs were between 0.97 and 1 unless there were 4 binary predictors or 16 binary predictors with minimum node size 20. The median discrimination loss, i.e. the difference between the median test AUC and the true AUC, was 0.025 (range 0.00 to 0.13). Median training AUCs had Spearman correlations of around 0.70 with discrimination loss. Median test AUCs were higher with higher events per variable, higher minimum node size, and binary predictors. Median training calibration slopes were always above 1, and were not correlated with median test slopes across scenarios (Spearman correlation -0.11). Median test slopes were higher with higher true AUC, higher minimum node size, and higher sample size.

**Conclusions**: Random forests learn local probability peaks that often yield near perfect training AUCs without strongly affecting AUCs on test data. When the aim is probability estimation, the simulation results go against the common recommendation to use fully grown trees in random forest models.

## 1 Background

Random Forests (RF) is an ensemble learning method introduced by Leo Breiman in 2001 (1). The difference between RF and other tree ensemble methods such as bagging or boosting is that the trees in RF are independent. A bootstrap sample is selected at each tree and at each node of each tree a random subset of predictors is considered for the best split. This reduces the correlation between trees.

---

[*] Corresponding author: Ben Van Calster e-mail: ben.vancalster@kuleuven.be

RF is used across a variety of clinical problems (2) and in recent years it has become very popular for clinical prediction modelling (3–6). Its popularity has risen due to its good reported performance in applied studies, and its claimed robustness against overfitting in combination with the limited need for hyperparameter tuning (7). It has been reported that RF models have better performance when the individual trees in the ensemble are overfitted (8–10). Although RF has been widely investigated as a 'classifier', but the literature about their performance as probability estimation trees (PET) is scarce.

In a recent study of women with an ovarian tumor, we compared the performance of different machine learning algorithms to estimate the probability of five tumor types (benign, borderline malignant, stage I primary invasive, stage II-IV primary invasive, secondary metastatic) (11). We developed prediction models on training data (n=5909) using multinomial logistic regression (MLR), ridge multinomial logistic regression, RF, XGBoost, neural networks, and support vector machines with Gaussian kernel. We evaluated discrimination performance using the Polytomous Discrimination Index (PDI) as a multiclass area under the receiver operating characteristic curve (AUC) (12,13). The PDI is the probability that, when presented with a random patient from each category, the model can correctly identify the patient from a randomly selected category. With five outcome categories, the PDI is 1/5 for an uninformative or random model and 1 for a model with perfect discrimination. We observed that RF had near perfect discrimination on the training data (PDI 0.93 for RF vs 0.47-0.70 for other models), and was competitive on the external validation data (PDI 0.54 for RF vs 0.41-0.55 for other models; n=3199) (**Table S1**). The observation that the RF model had near perfect (i.e. highly suspicious) discrimination on training data, yet performed competitively during external validation, may be somewhat counterintuitive. Such high training set performance suggests strong overfitting by modeling considerable amounts of noise, which would lead to reduced performance on new data(14,15). This was a striking observation for us: the training results suggest a suspiciously high degree of overfitting by RF compared to other models, such that we would have expected a stronger reduction in performance on new data for RF. In this study we aimed to understand the behavior of random forests for probability estimation by (1) visualizing data space in three real world case studies and (2) conducting a simulation study.

The paper outline is as follows. In Section 2 we summarize the RF algorithm for probability estimation, in section 3 we visualize the predictions for the ovarian tumor data, and present two additional case studies. In section 4, we present a simulation study to explore the effect of trees depth and training sample size and the data generation mechanism (DGM) to better understand the behavior of the RF algorithm. In Section 5 we discuss our findings.

## 2   Random forest for probability estimation

When the outcome is categorical, RF can be used for classification or for probability estimation. In this work we will use random forest for probability estimation (16,17). RF is a tree-based ensemble method, and when used for probability estimation it works as follows:

1. Draw *ntree* bootstrap samples from the original training dataset, where *ntree* denotes the number of trees in the forest.

2. On each bootstrap sample, construct a tree using recursive binary splits. To reduce correlation between trees, a number of predictors (*mtry*) is chosen randomly at each split. *mtry* is a hyperparameter and can be tuned, but often a default value equal to the square root of the total predictors (P) is used. A split on one of these predictors is chosen so that the selected splitting criterion (e.g. Gini index) is optimized.

3. Splits are consecutively created as long as all child nodes contain a specific minimum number of observations (*min.node.size*). When a node cannot be split without violating this condition, the node becomes a final leaf node. Other stopping criteria can be defined (18). For each leaf node, the proportion of cases from each outcome class can be calculated. Alternatively, the majority vote can be determined: the outcome class that has the most cases in the leaf.

4. To obtain a probability estimate for each outcome class *i* for a new case, we first determine the new case's appropriate leaf node for each of the *ntree* trees. Then, two basic approaches are possible. The first uses the proportion of the *ntree* majority votes (cf step 3) that equal *i*. The second averages the proportion of cases from class *i* (cf step 3) across the *ntree* trees (16).

In the seminal books "The Elements of Statistical Learning" and "An Introduction to Statistical Learning", the authors highlight the simplicity of training RF models (14,19). Regarding the commonly encountered claim that RF cannot overfit, the authors indicate that increasing *ntree* does not cause overfitting. It has been suggested that *ntree* does not need to be tuned, but that too low values lead to suboptimal performance (7,20). A value of 500 or even 250 has shown to be sufficient in most applications (7). A typical value for *mtry* is $\sqrt{P}$, as recommended by Breiman, or lower values to maximize decorrelation (14). Hastie and colleagues suggest that *min.node.size* can be set to a very low value, even 1, and that *mtry* is a more important tuning parameter: *"when the number of variables is large, but the fraction of relevant variables small, random forests are likely to perform poorly with small mtry. ... Our experience is that using full-grown trees seldom costs much, and results in one less tuning parameter"* (14).

## 3  Case studies

### 3.1  Methods

We aimed to visualize the estimated probabilities in a data space to obtain a better understanding of the phenomenon where RF models with near perfect discrimination also performed competitively during external validation. We followed a typical random train-test split used in machine learning procedures. We developed RF and MLR prediction models on the training set using two continuous and a number of categorical predictors. We use only two continuous predictors because if we set the categorical predictors to a fixed value, e.g. the most common one, we can show a two-dimensional subset of the complete data space by showing the two continuous predictors on the x-axis and y-axis. We can show estimated probabilities in this subset as a heatmap, and show individual cases (from training or test set) as a scatter plot. This allows us to visualize how RF and MLR transform predictor values into probability estimates, for example in terms of smoothness. Obviously, only cases for which the categorical values equal the chosen fixed value can be shown. By choosing different fixed values for categorical variables, we can visualize different subsets of data space. We noticed that the range of estimated probabilities was larger for RF than MLR. Therefore, to ensure a proper visualization of the high and low risk estimates, the greyscale in the heatmaps is bounded to the minimum and maximum predicted probabilities by each model in each panel. We also include figures using the same scale for all heatmaps in **Additional File 1**.

The RF models were trained with ranger package, with *ntree*=500, *mtry*= $\lceil\sqrt{P}\rceil$, and min.node.size=2. Ranger estimates the probabilities with Malley's probability machines methods which averages the proportion of cases from each class over the terminal nodes from each of the trees (16,21). In MLR models, we modelled continuous predictors using restricted cubic splines (rcs) with 3 knots to allow nonlinear associations (22,23). For each model we calculated the train and test PDI, and multinomial calibration plots. The code for training the models and generating the plots is available in the **OSF repository (https://osf.io/y5tqv/).**

### 3.2 Ovarian cancer diagnosis

This prospective study collected data of patients between 1999 and 2012. All patients had at least one adnexal (ovarian, para-ovarian, or tubal) mass that was judged not to be a physiological cyst, provided consent for transvaginal ultrasound examination, were not pregnant, and underwent surgical removal of the adnexal mass within 120 days after the ultrasound examination. We randomly split the data (N=8398) into training (n=5900, 70%) and test parts (n=2498, 30%), and developed models on the training data using patient age (in years) and CA125 (in IU/L) as continuous variables, and five ultrasound based categorical variables (proportion of solid tissue, number of papillary projections, if the mass has more than 10 locules, if the mass has shadows and if the mass has ascites). Note that the proportion of solid tissue is a continuous variable that can be seen as a semi-categorical variable with 75% of observations having values 0 or 1. The distribution of classes in the dataset was 66% (5524) for benign tumor, 6% (531) for borderline ovarian tumor, 6% (529) for stage I ovarian cancer, 17% (1434) for stage II-IV ovarian cancer and 5% (380) for metastatic cancer to the ovaries (detailed information in **Table S2**).The apparent PDI was 0.97 for RF and 0.52 for MLR. In the test set the PDI decreased to 0.56 for the RF model and remained 0.52 for the MLR.

**Figure 1** shows heatmaps for the estimated probabilities of a benign, borderline, stage I invasive, stage II-IV invasive, and secondary metastatic tumor, with training data cases superimposed (See **Figures S1-2** for extended visualizations). Cases belonging to the class to which the probabilities refer are shown in red, other cases in green. One set of heatmaps refer to the fitted RF model, the other to the fitted MLR model. Whereas estimated probabilities from the regression model change smoothly according to the values of the continuous predictors, the estimated probabilities from the RF model peak where events from the training data were located. Where many events were found in proximity, these peaks combined into a larger area with increased probability. For events in less densely populated areas of data space, these peaks were idiosyncratic: in test data, events in these areas of data space tend to be located in different places (**Figures 2 and S3-S4**). The calibration performance of the RF model was very poor in the training set: high probabilities were underestimated and low probabilities were overestimated (**Figure 3**). Calibration in the test set was much better.

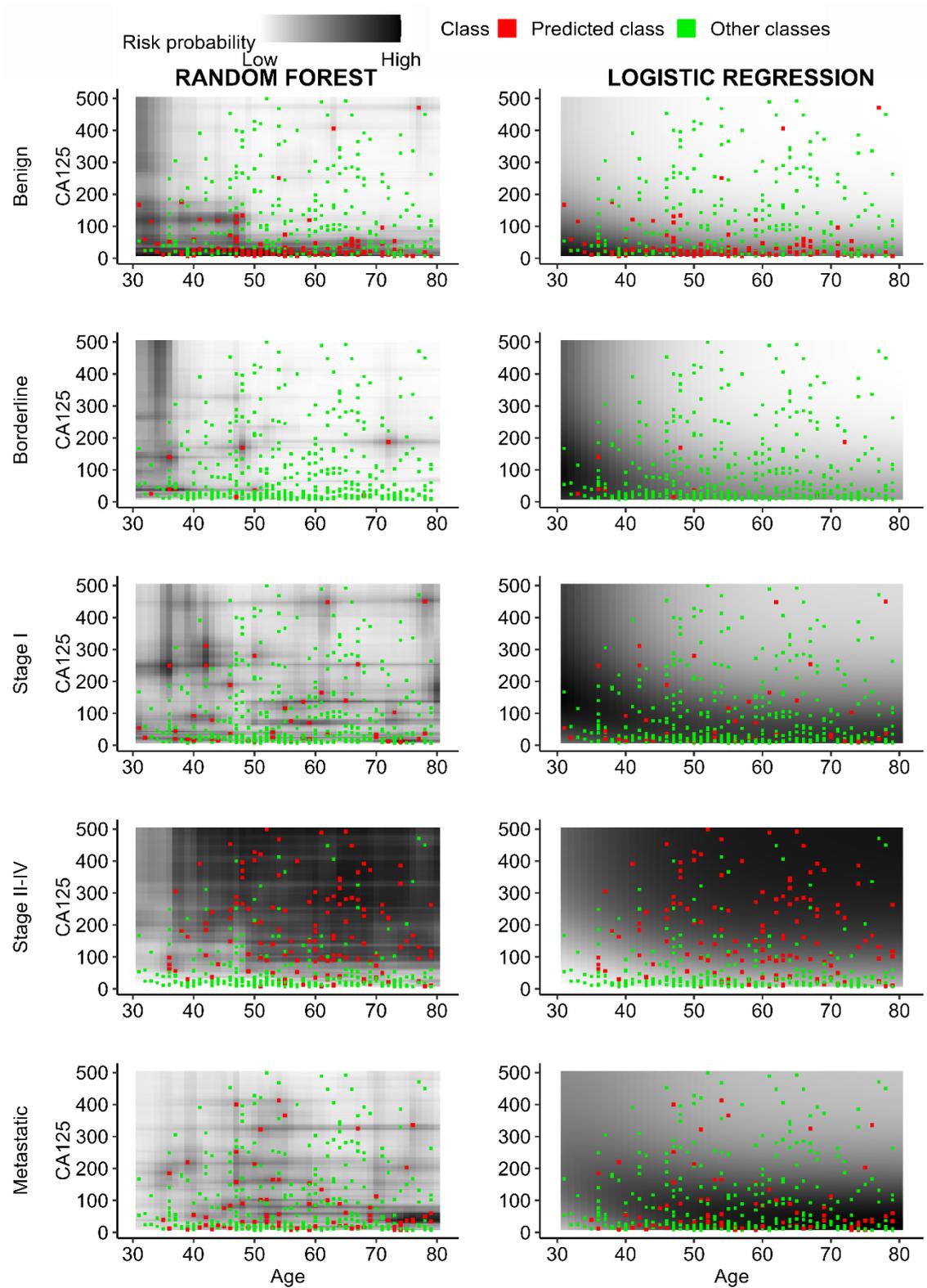

**Figure 1** Random Forest and logistic regression probability estimation in data space for 4 subtypes of ovarian malignancy diagnosis with **cases in training set superimposed**. CA125 bounded to 500.

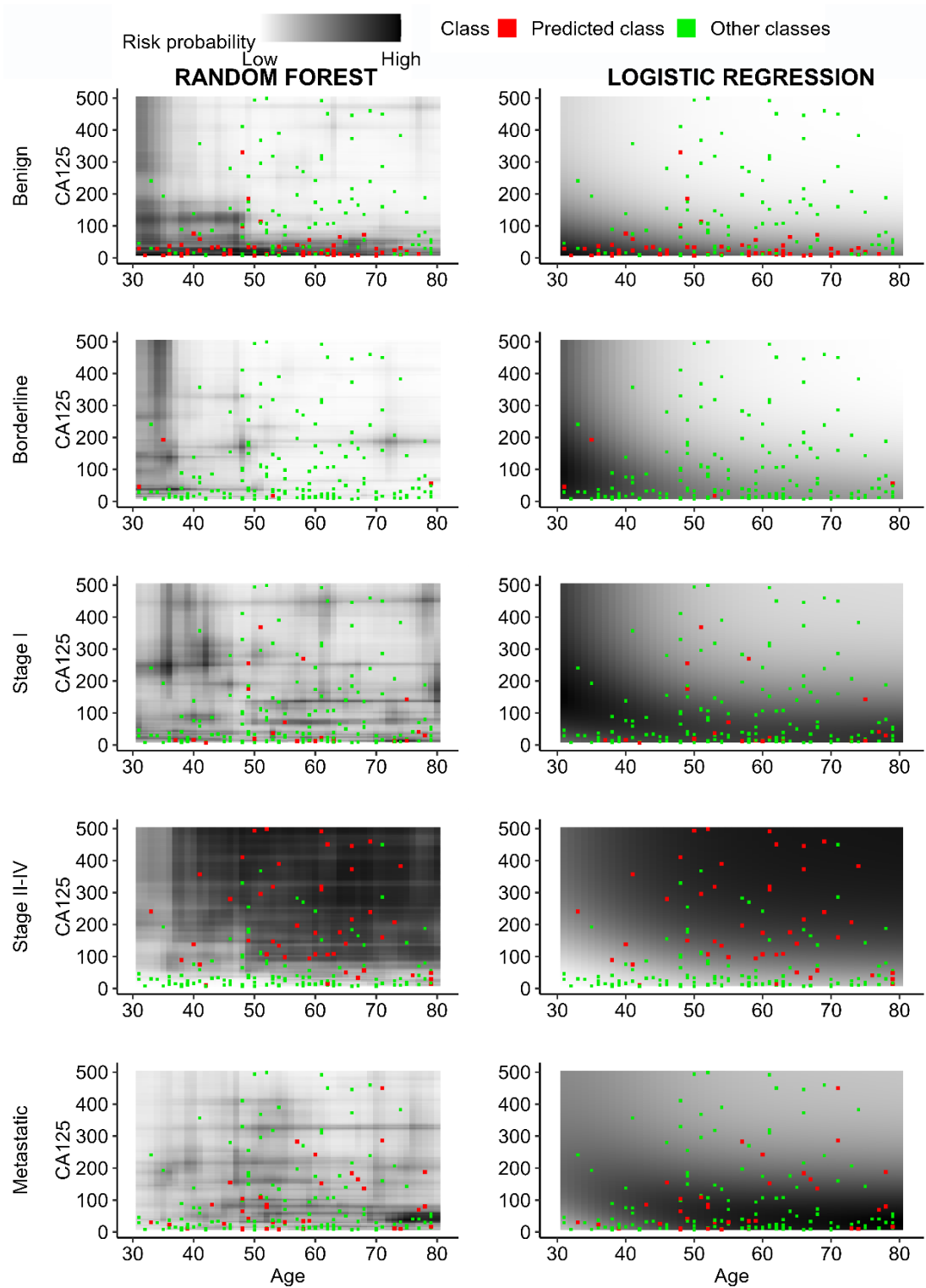

**Figure 2** Random forest and logistic regression probability estimation in data space for 4 subtypes of ovarian malignancy diagnosis with **cases in test set superimposed**. CA125 bounded to 500.

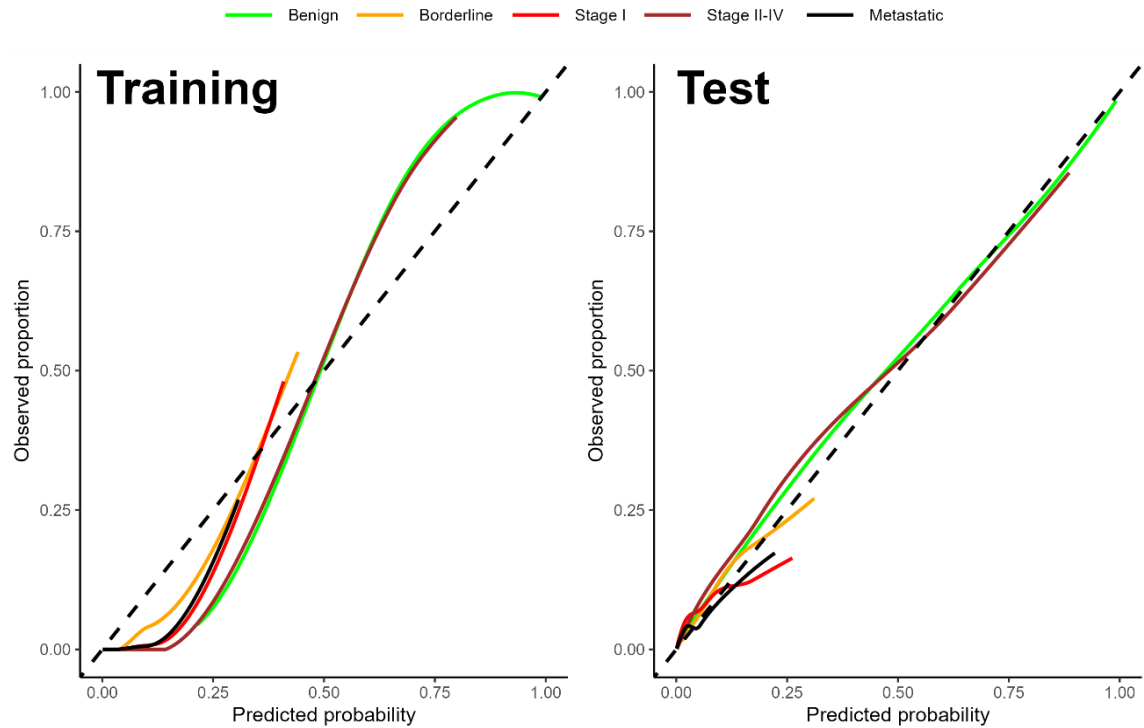

**Figure 3** Calibration plots for random forest model in ovarian cancer data. Observed proration is estimated with a LOESS model. The plots only show observed proportions for predicted probabilities between quantiles 5th and 95th.

### 3.3 CRASH3: traumatic brain injury prognosis

CRASH-3 data was collected between 2012 and 2019 for a multicenter, randomized, placebo-controlled trial to measure the effects of tranexamic acid on death, disability, vascular occlusive events and other morbidities in 12660 patients with acute traumatic brain injury (TBI) (24). We used age (years) and systolic blood pressure (mmHg) as continuous variables, and sex, Glasgow Coma Scale (GCS) eye-opening (4 levels), and pupillary reaction (4 levels) as categorical variables. We perform a complete case analysis (CCA) removing patients for which one or more values were missing obtaining a complete dataset of 12548 patients. CCA was used for simplicity and because the phenomenon under study should not be affected importantly by this. The outcome was measured 28 days after randomization: alive (n=10022, 80%), death due to head injury (n=2309, 18%), or death other cause (n=217, 2%) (detailed information in **Table S3**). The training set included 8783 patients (70%), the test set 3765 (30%).

For RF, the PDI was 0.96 in train data and 0.54 in test data. For MLR, the PDI was 0.61 and 0.60, respectively. The heatmaps drew a similar picture compared with the previous case study: RF had clear probability peaks whilst MLR had smoothly changing probabilities (**Figures S5-8**)The calibration plots for RF were also similar: poor calibration in training data but decent in the test data for the 2 most common outcomes (**Figure S9**).

### 3.4 IST: Type of stroke diagnosis

The International Stroke Trial (IST) database was designed with the aim of establishing whether early administration of aspirin, heparin, both or neither influenced the clinical course of acute ischaemic stroke (25). Data for the 19435 patients with suspected acute ischaemic stroke were recruited between 1992 and 1996. We use age (years) and systolic blood pressure (mmHg) as continuous variables, and conscious state (fully alert vs drowsy) , deficit of face (yes vs no), deficit of arm/hand (yes vs no), deficit of leg/foot (yes vs no), dysphasia (yes vs no), and hemianopia (yes vs no) as categorical variables. We again performed a CCA retaining 15141 patients. The outcome is the type of stroke: ischaemic (n= 13622, 90%), indeterminate (n= 736, 5%), haemorrhagic (n= 439, 3%), or no stroke (n= 344, 2%) (detailed information in **Table S4**). The training set included 10598 patients (70%), the test set 4543 (30%).

The RF model had a training PDI of 0.89 and a test PDI of 0.35. For MLR, the training set PDI was 0.39, the test set PDI 0.41. In this dataspace the phenomenon is notorious, with very local peaks around train cases (**Figures S10-13**). The calibration plots for training and test data showed poor calibration (**Figure S14**).

## 4 Simulation study

### 4.1 Aim

We conducted a simulation study to assess which key factors of the modelling setup (dataset and minimum node size) contribute to phenomenon of having an exaggerated AUC in the training data without strong signs of overfitting in test data. We report the simulation study using the ADEMP (aims, data-generating mechanisms, estimands, methods, and performance measures) structure (26). The code for the simulation study can be found in the OSF repository (https://osf.io/y5tqv/).

### 4.2 Data Generating Mechanism (DGM)

For the simulation study, we generated data by assuming that the true model was of the form of a MLR with an outcome event fraction of 0.2 (see **Additional File 1: A3 Simulation Algorithm** for details).

The 48 DGMs differed according to the following parameters:

i. **Predictor distribution**: predictors were either all continuous with multivariate normal distribution or all binary with 50% prevalence.

ii. **Number of predictors**: there were either 4 true predictors (0 noise predictors), 16 true predictors (0 noise predictors), or 16 predictors of which 12 noise predictors. Noise predictors had a true regression coefficient of 0.

iii. **Correlation between predictors**: Pearson correlations between all predictors were either 0 or 0.4.

iv. **True AUC:** this was either 0.75 or 0.90.

v. **Balance of regression coefficients:** the true model coefficients that were not 0 were either all the same (balanced) or not (imbalanced). When imbalanced, one fourth of the predictors have a coefficient that is 4 times larger than the others.

The true model coefficients for generating the data were obtained by trial error and are available in **Table S5**.

### 4.2.1 Estimands

For both training and test data, we estimate model discrimination, whether risk estimates have too high (overconfidence) or too low (underconfidence) spread, and the prediction error. Underconfidence reflects the situation in Figure 3 (Train): high probabilities are underestimated, low probabilities are overestimated. Overconfidence is the opposite: high probabilities are overestimated, low probabilities are underestimated. For test data, we also calculate discrimination loss vs the true model, and the relative contribution of bias and variance to the prediction error. As the training and test samples are based on the same DGM, the test results reflect internal rather than external validation.

## 4.3 Methods

We fitted models on training datasets of size 200 (small) or 4000 (large), and for RF models we used values for *min.node.size* of 2 or 20 and ranger package for training. As a result, there were 2x2x3x2x2x2x2 = 192 scenarios. For each scenario, 1000 simulation runs were performed (i.e. 1000 different training datasets). For RF, *ntree* was fixed at 500, and *mtry* at the square root of the number of predictors (default value). Models were validated on a single large test dataset per DGM (N=100.000) to avoid sampling variability.

## 4.4 Performance

For discrimination we calculated the AUC and for confidence of risk estimates the calibration slope (slope <1 means overconfidence, slope >1 underconfidence). Calibration slope is calculated as the slope of a logistic regression (LR) model fitting the outcome to the logit of the estimated probabilities as the only predictor. Calibration intercept is calculated fitting the same model for a slope of 1 by setting the predicted probabilities as an offset term. Discrimination loss was calculated as the difference between true AUC and median test AUC. Finally, the mean squared error (MSE) of the predicted probabilities was calculated according to (27) as the sum of squared bias and variance (see **Additional File 1: A4 Simulation metrics** for details).

## 4.5 Results

The aggregated simulation results using median and interquartile range for discrimination and calibration and mean and standard deviation for mean squared error are available in **additional file 1** (**Table S6**). The complete simulation including the code and 1000 simulation for each of the 192 scenarios are available in the OSF repository (https://osf.io/y5tqv/).

### 4.5.1 Discrimination

In the simulation study the median training AUCs were close to 1 in most of the cases. The median training AUC was between 0.97 and 1 unless there were 4 binary predictors, or 16 binary predictors combined with a minimum node size of 20 (**Figure 4**). Higher *min.node.size* resulted in less extreme training AUCs.

In general, median test AUCs were higher when there was a large vs small training dataset, high vs low *min.node.size*, high vs low correlation between predictors, binary versus continuous predictors, and 4 versus 16 predictors (except with correlated continuous predictors) (**Figures 5**). All other simulation factors being equal, the scenarios with 4 true and 12 noise predictors had results for the AUC that were identical (**Figures 4-5 and S15-16**).

The Spearman correlation between median training AUC and discrimination loss was 0.72 for scenarios with true AUC of 0.9, and 0.69 for scenarios with true AUC of 0.75 (**Figure S17**). The median discrimination loss was 0.025 (range 0.00 to 0.13). In the 114 scenarios where the median training AUC was ≥0.99, the median discrimination loss was 0.036 (range 0.003 to 0.13). In the other scenarios, the median discrimination loss was 0.013 (0.00 to 0.069).

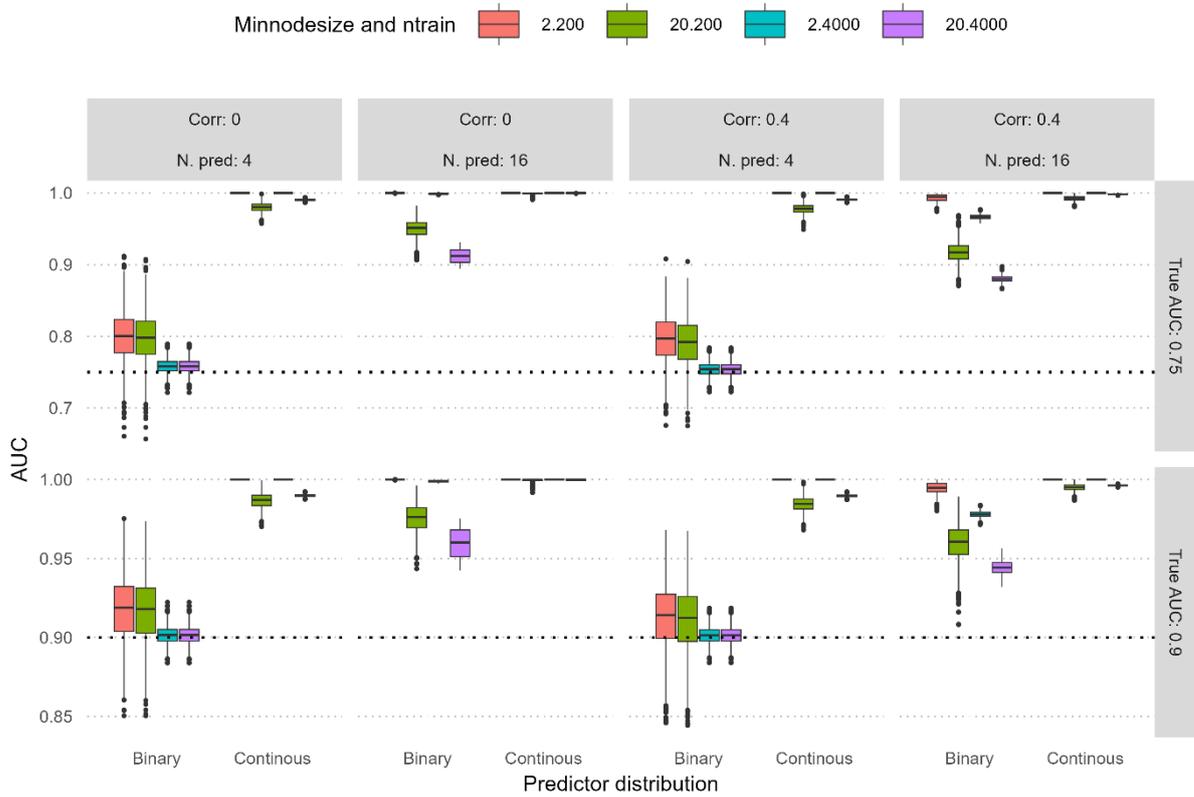

**Figure 4** Training AUC by simulation factors and modelling hyperparameters in scenarios without noise. Scenarios are aggregated by strength because this simulation factor had minimal effect.

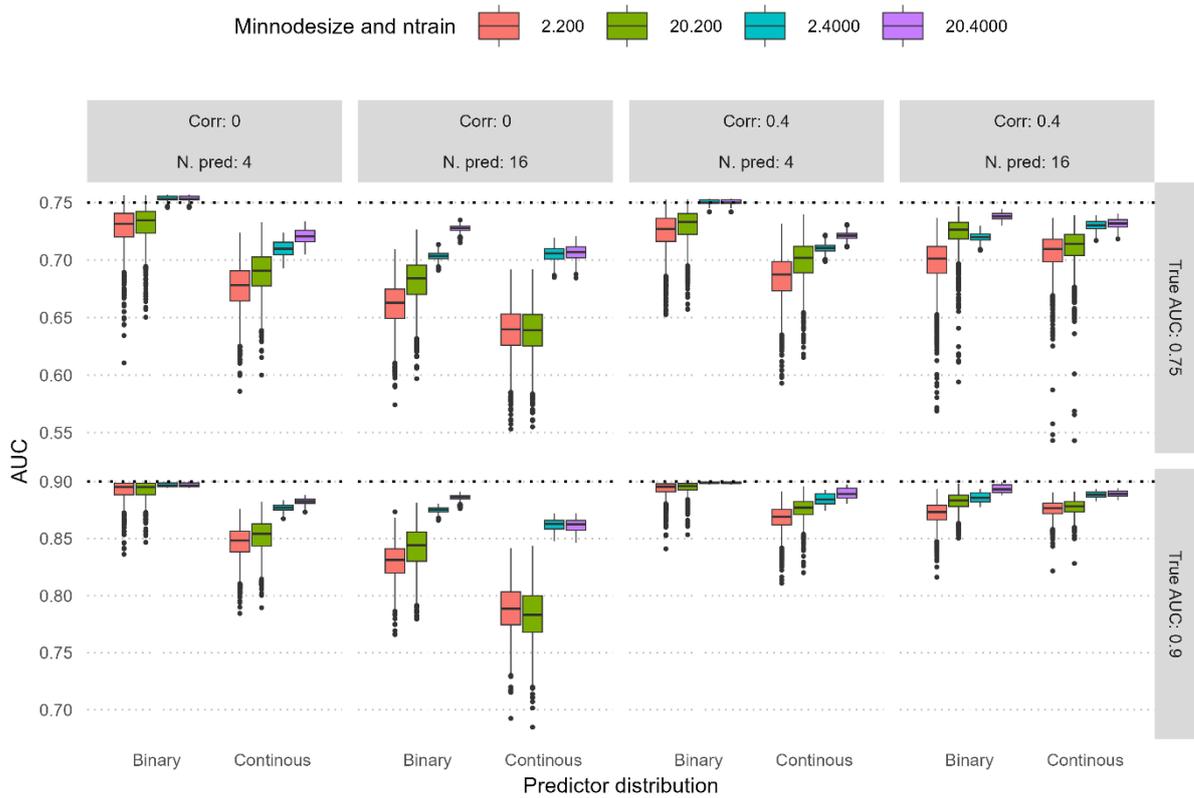

**Figure 5** Test AUC by simulation factors and modelling hyperparameters in scenarios without noise. Scenarios are aggregated by strength.

#### 4.5.2 Calibration

Median training calibration slopes ranged between 1.10 and 19.4 (**Figure 6 and S18**): the probability estimates were always underconfident where high probabilities were underestimated and low probabilities overestimated. This is the consequence of perfect separation between events and non-events in training data (i.e AUC =1) which means that any estimation above 0 or below 1 is underconfident (**Figure S19**). The median slope was lowest in scenarios with few binary predictors or higher min.node.size. Median test calibration slopes ranged between 0.45 and 2.34. Across all scenarios, the Spearman correlation between median training slope and median test slope was -0.11 (**Figure S17**). Median test slopes were mainly higher when the true AUC or min.node.size were higher. In addition, median test slopes tended to be higher with binary predictors, uncorrelated predictors, and higher sample size (**Figure 7 and S18**). Calibration slopes were similar in scenarios with 16 true predictors and 4 true predictors and 12 noise predictors (**Figures 6-7 and S20-21**). In the 78 scenarios without perfect training (AUC <0.99) the median test calibration slope was between 0.59 and 2.34 with median of 1.10. In the 114 scenarios with almost perfect training AUC (≥0.99) the median calibration slope was 0.92.

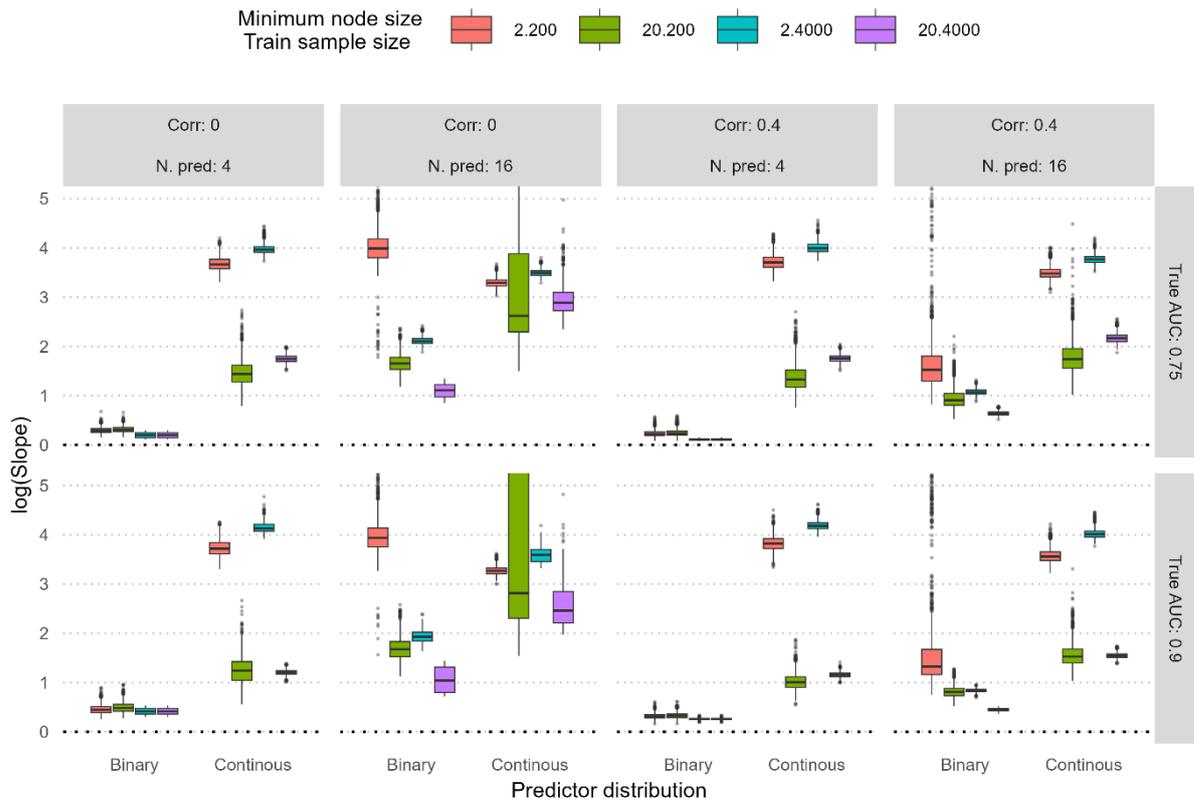

**Figure 6** Training set calibration log slope by simulation factors and modelling hyperparameters in scenarios without noise. Scenarios are aggregated by strength. The ideal value for log slope is 0.

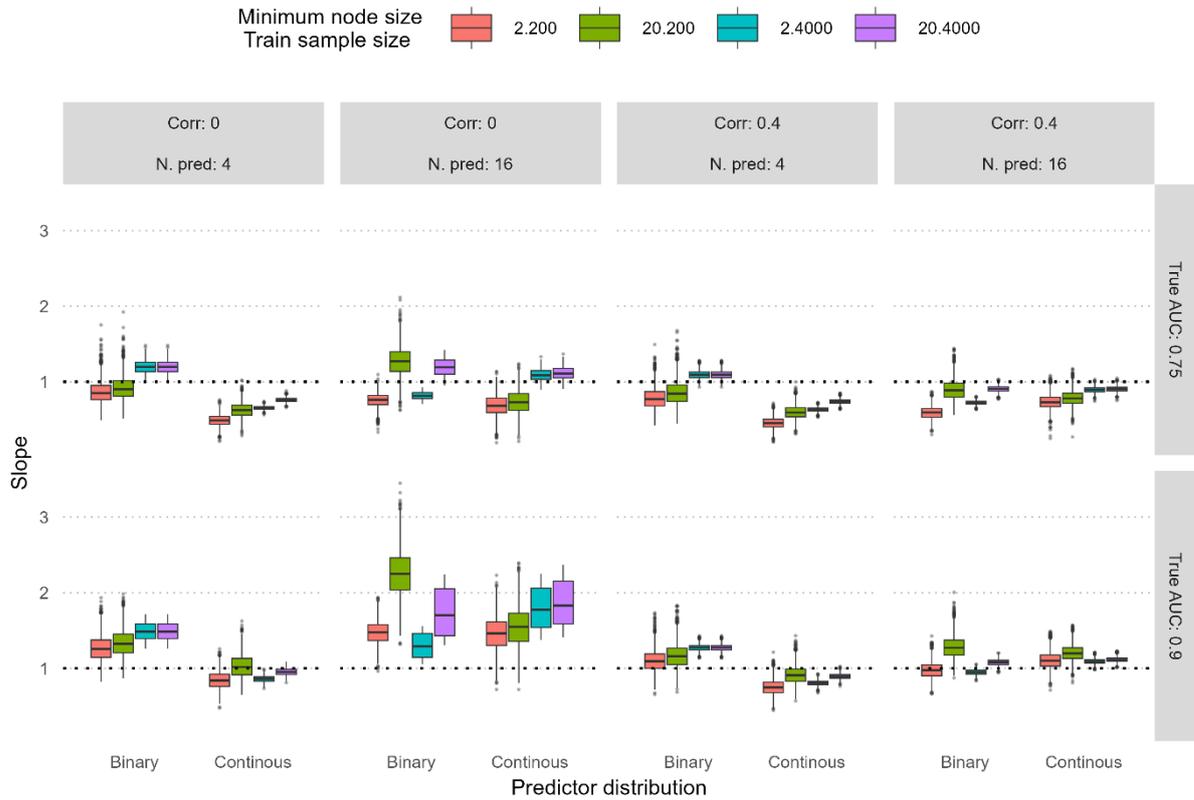

**Figure 7** Test set calibration slope by simulation factors and modelling hyperparameters in scenarios without noise. Scenarios are aggregated by strength. The ideal value for the slope is 1.

### 4.5.3 Mean squared error (MSE)

Median MSE across scenarios was 0.008 (range 0.000-0.045) with median squared bias of 0.002 (range 0.000-0.038) and median variance of 0.005 (range 0.000-0.017). For the 114 scenarios with median training AUC ≥0.99, we observed a median test MSE of 0.010 with a median squared bias of 0.004 (range 0.0004-0.0384) and a median variance of 0.006 (range 0.001-0.017). For the rest of the scenarios, the median test MSE was 0.006. Across all scenarios, the Spearman correlation of mean test squared bias and mean test variance with median training AUC were 0.47 and 0.43, respectively. The correlation with the discrimination loss was of 0.51 for the squared bias and 0.70 for the variance. Lower sample size in training was associated with higher median test variance, and with higher median test squared bias in scenarios with continuous predictors. Lower min.node.size (i.e. deeper trees) was associated with lower variance but higher bias in test data when training sample size was small. More predictors were associated with higher bias, whereas correlation between predictors and higher true AUC were associated with lower bias **(Figure 8).** The models with noise predictors had lower variance and higher bias compared to scenarios with 4 true and no noise predictors (**Figure S22**).

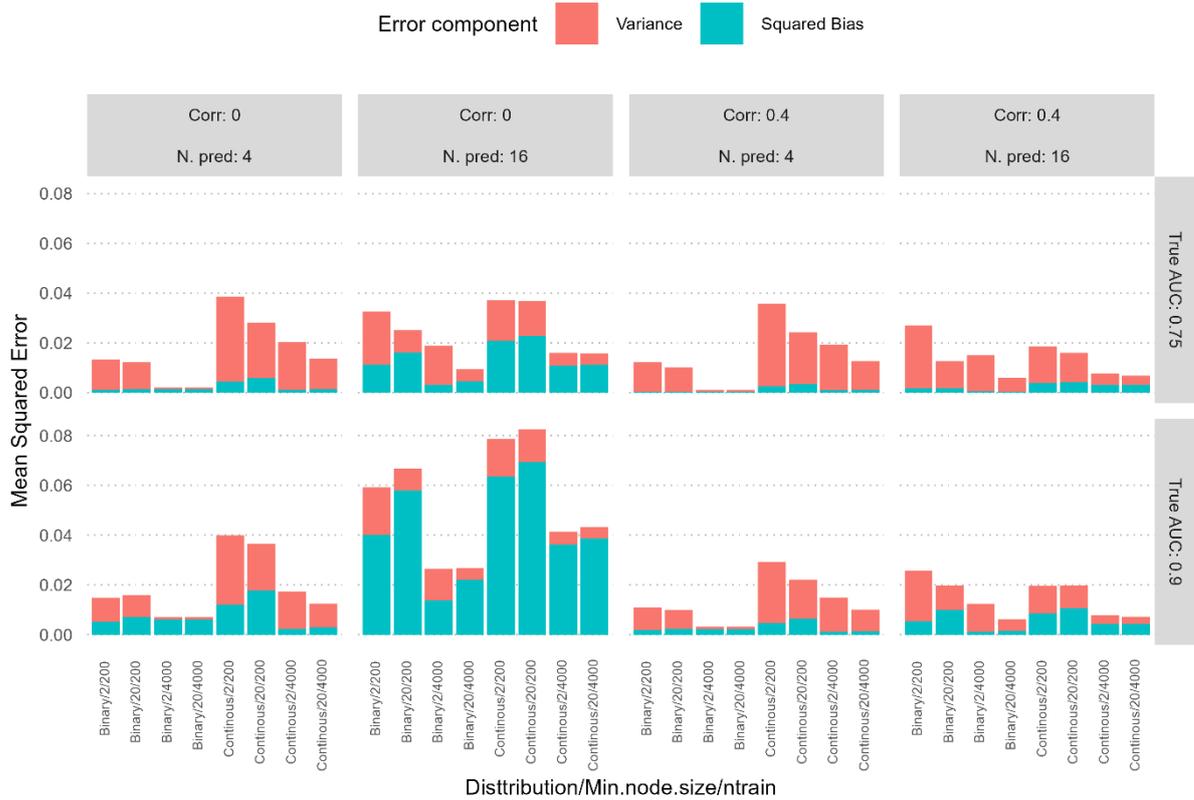

**Figure 8**    Mean squared error across scenarios without noise aggregated by strength.

## 5    Overall discussion

We tried to better understand and visualize the behavior of random forests for probability estimation. We make three key observations from this work. First, RF models learn local probability peaks around training set events, in particular when the trees are very deep (i.e. low *min.node.size*) and when there are continuous predictors. Where a group of events are located close to one another in 'data space', the probability peaks are combined into a region of increased probability. Where events are isolated in data space, the probability peaks are very local. Learning through peaks leads to very optimistic (often near perfect) discrimination in training data, but also to reduced discrimination in new data compared to models that rely on less deep trees (cf. simulation settings where RF models with *min.node.size* 2 vs 20 were fitted on a set of continuous predictors). Probably, the reduction in discrimination on new data is modest because the local peaks for isolated events are often harmless for new data: it is unlikely to see an event in the exact same location. Second, RF also suffers from "classical overfitting" in which models with higher discrimination on training data tend to have lower discrimination on new data than models with less optimistic discrimination (cf. simulation settings where RF models are learned on 4 binary predictors using 200 vs 4000 training cases). Third, in training and test data, calibration performance for RF models is different from what we commonly observe for LR models. Whereas LR based on maximum likelihood leads by definition to calibration slopes of 1 on training data, calibration slopes for RF models were always above 1 on training data. Also, as opposed to LR, calibration slopes for RF models do not converge to 1 on new data. This different behavior is probably caused by the pragmatic way in which probabilities are obtained for RF models, whilst LR estimates probabilities in a principled way through maximum likelihood.

The simulation results regarding discrimination and calibration go against fitting very deep trees when using RF for probability estimation. This is in line with recent work that illustrated that RF using deeply grown trees results in risk estimates that are particularly unstable (28). The heatmaps for our case studies illustrate how RF models with deeply grown trees lead to probabilities that change non-smoothly with changes in the values of predictors. It has been suggested to set *min.node.size* to 5% or 10% of the sample size (16,29). Alternatively, *min.node.size* can be tuned. Although this was not the focus of our study, it seems natural to tune with logloss (also known as the negative loglikelihood or cross-entropy) or Brier score as the loss function since they capture calibration (18). In their work, Ledger and colleagues tuned *min.node.size* by optimizing logloss based on 10-fold cross-validation for 30 random values for mtry and *min.node.size* using the trainControl and train functions from the caret R package (11). This resulted in *mtry*=3 and *min.node.size*=15, with competitive results in test data. Applying the tuneRanger R package with 200 iterations for our case studies based on logloss yielded an optimal *min.node.size* of 8 (0.1% of training set size) for ovarian cancer data, 261 (2.1%) for CRASH3 data and 591 (3.9%) for IST data (18). These values are higher than the default values in many statistical software programs.

Three comments regarding the interpretation of discrimination and calibration results for RF models are worth making. First, it is well known that apparent performance, i.e. performance assessment on the exact same dataset that was used to train the model, is overly optimistic (30,31). Our work indicates that this is a fortiori the case for RF. Unfortunately, some studies present their RF models with "excellent" performance because they only present discrimination for the training data (4,32,33). Instead, proper internal and external validation results should be reported. Second, from a regression modeling perspective, a calibration slope that is clearly below 1 on internal validation is a symptom of overfitting. This cannot be applied in the same way for RF models. Models based on larger training samples had higher calibration slopes in our simulation study, but a calibration slope of 1 does not appear to have a special meaning. Models with a calibration slope above 1 on test data when trained on 200 training samples, had an even higher slope when trained on 4000 samples. The interpretation of risk estimates based on RF requires caution, probably because probabilities are generated in a very ad hoc way. Of course, the calibration slope still quantifies in a descriptive manner whether the risk estimates are on average too confident (slope < 1), or not confident enough (slope > 1), or fine (slope = 1). Third, despite that RF models had high discrimination results in the training data (suggestive of overfitting), the calibration slope in the training data was always above 1 (risk estimates show too little spread, suggestive of underfitting). This appears to be a consequence of the bootstrapping procedure in combination with the low *min.node.size*. Due to the bootstrapping, a training set case was part of approximately 63% of bootstrap samples, and therefore was used for 63% of the trees in the forest. When averaging the proportion of events in the appropriate leaf nodes over all *ntree* trees to get a probability estimate for a given training set case, 63% of these proportions are near perfect: close to 1 if the training case is an event, close to 0 if the training cases is a non-event. The remaining 37% are more variable and often far less good. The 63% near perfect proportions cause discrimination to be very high, because most events will end up with an estimated probability of event that is higher than that for most non-events. The remaining 37% of the proportions pull the probability estimates away from 0 (if the case is a non-event) or 1 (if the case is an event), leading to calibration slopes > 1.

Although the aim of our paper was largely educational, it links to previous more fundamental work and fills a gap in the literature by explicitly studying factors that contribute to better discrimination and calibration on new data. Wyner and colleagues (2017) argued that RF has excellent performance because it is an "interpolating classifier", i.e. it is fitted with little to no error to the train data (34). They argue that the interpolation should not be confused with overfitting. Even if the individual trees are overfit, each training set case is not used in about 37% of the individual trees, such that averaging over trees partially solves overfitting. Belkin and colleagues have linked this to a double descent curve for highly flexible algorithms: when the complexity of the model increases, test set performance first improves, then deteriorates, and finally improves again once the 'interpolation threshold' (where perfect training performance is achieved) is exceeded (35). Recently, however, Buschjäger and Morik opposed to the existence of double descent in RF (36). Mentch and Zhou linked the success of RF to the signal to noise ratio (SNR) of the data. In their work they present that randomness of RF is beneficial in situations with low SNR whereas bagging is preferred if SNR is high (37). They explain the success of RF by the low SNR of many real world datasets. This view contradicts with the view that flexible algorithms work best when the SNR is high (38). Finally, the issue of calibration of estimated probabilities in the context of RF models received little attention in the literature, although it is key for optimal clinical decision making (39).It has been suggested that using fully grown trees in RF leads to suboptimal risk estimates (16,29). However, this is rarely mentioned and hence it is common to see that low minimum node sizes are recommended because the problem is treated as a classification problem instead of probability estimation problem (e.g., see default for randomForest package or scikit-learn). We think that the current study sheds further light on probability estimation in RF. Of course, a generic alternative that works for any miscalibrated model is to recalibrate the probabilities of the RF afterwards using new data (40).

We identified the following limitations of our study. Firstly, a simulation study is always limited by the included scenarios. It would be of interest to include more simulation factors and values per simulation factor (e.g. for *min.node.size*) in the simulation study, or to include scenarios where RF hyperparameters are tuned rather than fixed. Tuning could improve the calibration of the models (18,41). However, the simulation study already had 192 scenarios and adding more factors or values would increase the computational cost exponentially and would difficult the interpretation of the results. Topics that could be investigated in further simulation studies include varying other hyperparameters than *min.node.soze* (e.g. *mtry*, sampling fraction, splitting rule) and investigating more values for sample size, number of predictors, proportion of noise predictors, and min.node.size. Secondly, we were using logistic DGMs without nonlinear or nonadditive associations with the outcome. We assumed that the impact of this would be limited, because the focus was not on the comparison of RF with LR, and any nonlinearity or nonadditivity (including the absence of it) has to be learned by the algorithm. Thirdly, the traditional RF algorithm selects variables at each split in a way that favors continuous over binary variables (42). Continuous variables can split in many ways, such that there is often a split that, perhaps by chance, has a better Gini impurity reduction than the Gini impurity reduction for a binary variable. The splits for continuous variables may often overfit, thereby increasing training discrimination but decreasing test discrimination. It is well documented that this affects variable importance measures (42), but it may also be relevant for model performance. The problem can be addressed by using an adapted RF algorithm such as cforest from partykit package (43). These adapted algorithms grow conditional inference trees (CIT) instead of classification trees. For the case studies, we observed that tuning and using the adapted RF yields a less optimistic training AUC and similar or slightly better test performance (See **OSF Repository, https://osf.io/y5tqv/**). However, our aim was to understand how the characteristics of the data and the modelling process affected the models, hence we did not systematically explore the effects of tuning or alternative algorithms.

We conclude that RF tends to exhibit local overfitting by learning probability peaks, in particular when the RF model is based on deeply grown trees. This local overfitting can lead to highly optimistic (near perfect) discrimination on the training data, but to reduced discrimination on new data compared to RF models based on less deeply grown trees. In line with the work of Kruppa and colleagues (41), our results go against the recommendation to use fully grown trees when using RF for probability estimation.

**List of abbreviations:**

RF: Random forest

PET: Probability estimation tree

MLR: Multinomial logistic regression

PDI: Polytomous discrimination index

AUC: Area under the receiver operating characteristic curve

TBI: Traumatic brain injury

GSC: Glasgow coma score

CCA: Complete case analysis

IST: International stroke trial

ADEMP: Aims, data-generating mechanisms, estimands, methods, and performance measures

DGM: Data generating mechanism

LR: Logistic regression

MSE: Mean squared error

SNR: Signal to noise ratio

## Declarations

**Ethics approval and consent to participate**   Not applicable

**Consent for publication**   Not applicable

**Availability of data and materials:**  All data and code that support the findings of this study are openly available in OSF repository at (https://osf.io/y5tqv/)(44) except the ovarian cancer dataset which is not publicly available.

**Competing interest**
*The authors have declared no conflict of interest.*

**Funding**            This research was supported by the Research Foundation – Flanders (FWO) under grant G097322N with BVC and DT as supervisors.

**Acknowledgements**   Partial findings from this paper where presented in Young Statisticians Meeting  2023 in Leicester (July)  and in the Conference of the International  Society of Clinical Biostatistics (ISCB) 2023 in Milan

(August).

# Additional File

## Overview Table for Additional File

| | |
|---|---|
| **Table S1** | Train and test performance of different machine learning algorithms on predicting ovarian type of tumour in terms of polytomous discrimination index (PDI) and ovarian malignancy in terms of AUC (Benign vs Malignant). AUC, Area Under receiver operating curve; MLR, Multiple Linear Regression; RF, Random Forest; XGBoost, Extreme gradient boosting; NN, Neural network; SVM, Support Vector Machine. |
| **Table S2** | Distribution of different classes in the ovarian cancer dataset. |
| **Table S3** | Distribution of different classes in the ovarian CRASH dataset. |
| **Table S4** | Distribution of different classes in the ovarian IST dataset. |
| **Table S5** | Coefficients and simulation factors |
| **Table S6** | Main simulation results |
| **Figure S1** | Random forest probability estimation in data space for ovarian malignancy diagnosis with random forest (left) and multinomial logistic regression (right). Squares refer to train cases. |
| **Figure S2** | Random forest probability estimation in data space for ovarian malignancy diagnosis with random forest (left) and multinomial logistic regression (right) using same scale for all panels. Squares refer to train cases. |
| **Figure S3** | Random forest probability estimation in data space for ovarian malignancy diagnosis with random forest (left) and multinomial logistic regression (right). Squares refer to test cases. |
| **Figure S4** | Random forest probability estimation in data space for ovarian malignancy diagnosis with random forest (left) and multinomial logistic regression (right) using same scale for all panels. Squares refer to test cases. |
| **Figure S5** | Random forest probability estimation in data space for 3 possible outcomes in CRASH 3 dataset. Squares refer to train cases. |
| **Figure S6** | Random forest probability estimation in data space for 3 possible outcomes in CRASH 3 dataset using same scale for all panels. Squares refer to train cases. |
| **Figure S7** | Random forest probability estimation in data space for 3 possible outcomes in CRASH 3 dataset. Squares refer to test cases. |
| **Figure S8** | Random forest probability estimation in data space for 3 possible outcomes in CRASH 3 dataset. Squares refer to test cases using same scale for all panels. |
| **Figure S9** | Multinomial calibration plots of CRASH training and test data. Observed proration is estimated with a LOESS model. The plot shows only predicted probabilities between quantiles 5th and 95th. |
| **Figure S10** | Random forest probability estimation in data space for 4 possible type of strokes in IST dataset. Squares refer to training cases. |
| **Figure S11** | Random forest probability estimation in data space for 4 possible type of strokes in IST dataset. Squares refer to training cases using same scale for all panels. |
| **Figure S12** | Random forest probability estimation in data space for 4 possible type of strokes in IST dataset. Squares refer to test cases. |
| **Figure S13** | Random forest probability estimation in data space for 4 possible type of strokes in IST dataset using same scale for all panels. Squares refer to test cases. |
| **Figure S14** | Calibration plot in training for IST dataset. Observed proration is estimated with a LOESS model. The plot shows only predicted probabilities between quantiles 5th and 95th. |
| **Figure S15** | Training AUC by simulation factors and modelling hyperparameters in scenarios with noise. Scenarios are aggregated by strength. |
| **Figure S16** | Test AUC by simulation factors and modelling hyperparameters in scenarios with noise. Scenarios are aggregated by strength. |
| **Figure S17** | Spearman correlations of principal metrics across all scenarios, scenarios with true AUCs 0.9 and scenarios with true AUC 0.75. |
| **Figure S18** | Train and test calibration log(slope) in scenarios without noise. Scenarios are summarised by simulation factors that had minor effect. |
| **Figure S19** | Histogram of predicted probabilities in training for different simulation scenarios. Minimum node size was always 2 and training sample size 4000 |
| **Figure S20** | Training set calibration log slope by simulation factors and modelling hyperparameters in scenarios with noise. Scenarios are aggregated by strength. Perfect calibration is 0. |
| **Figure S21** | Test set calibration slope by simulation factors and modelling hyperparameters in scenarios with noise. Scenarios are aggregated by strength. Perfect calibration is 1. |
| **Figure S22** | Mean squared error across scenarios with noise aggregated by strength. |
| **Appendix A3** | Metrics explained |
| **Appendix A4** | Simulation Algorithm |

## A.1 Tables

**Table S1** Train and test performance of different machine learning algorithms on predicting ovarian type of tumour in terms of polytomous discrimination index (PDI) and ovarian malignancy in terms of AUC (Benign vs Malignant). AUC, Area Under receiver operating curve; MLR, Multiple Linear Regression; RF, Random Forest; XGBoost, Extreme gradient boosting; NN, Neural network; SVM, Support Vector Machine.

| Model | PDI (95% CI) | | AUC (95% CI) | |
|---|---|---|---|---|
| | *Models with CA125* | | | |
| | Train | Test | Train | Test |
| MLR | 0.58 (056;0.60) | 0.54 (0.50;0.59) | 0.94 (0.93;0.95) | 0.92 (0.90;0.93) |
| Ridge MLR | 0.53 (0.52;0.55) | 0.49 (0.46;0.53) | 0.93 (0.92;0.95) | 0.90 (0.88:0.92) |
| RF | 0.93 (0.92;0.94) | 0.54 (0.50;0.59) | 0.99 (0.98;0.99) | 0.92 (0.90;0.94) |
| XGBoost | 0.70 (0.68;0.72) | 0.55 (0.51;0.60) | 0.96 (0.95;0.97) | 0.92 (0.90;0.94) |
| NN | 0.60 (0.58;0.62) | 0.54 (0.50;0.58) | 0.95 (0.93;0.96) | 0.92 (0.90;0.93) |
| SVM | 0.47 (0.46;0.49) | 0.41 (0.39;0.43) | 0.93 (0.92;0.94) | 0.89 (0.88;0.91) |
| | *Models without CA125* | | | |
| | Train | Test | Train | Test |
| MLR | 0.53 (0.51;0.54) | 0.51 (0.47;0.54) | 0.93 (0.92;0.94) | 0.91 (0.89;0.93) |
| Ridge MLR | 0.50 (0.49;0.52) | 0.47 (0.44;0.49) | 0.92 (0.91;0.94) | 0.89 (0.87;0.91) |
| RF | 0.83 (0.82;0.84) | 0.50 (0.46;0.54) | 0.97 (0.96;0.98) | 0.91 (0.89;0.93) |
| XGBoost | 0.62 (0.60;0.63) | 0.50 (0.46;0.54) | 0.94 (0.93;0.96) | 0.91 (0.89;0.93) |
| NN | 0.57 (0.55;0.58) | 0.50 (0.46;0.54) | 0.94 (0.93;0.95) | 0.91 (0.89;0.93) |
| SVM | 0.48 (0.46;0.49) | 0.42 (0.39;0.45) | 0.92 (0.91;0.94) | 0.89 (0.87;0.90) |

**Table S2** Distribution of different classes in the IOTA dataset.

| CLASS | ALL | TRAIN | TEST | IN FIGURES 1 AND S1-S2 (TRAIN)[A] | IN FIGURES 2 AND S3-S4 (TEST)[A] |
|---|---|---|---|---|---|
| BENIGN | 5524 | 3875 | 1649 | 154 | 72 |
| BORDERLINE | 531 | 359 | 172 | 13 | 4 |
| STAGE I | 529 | 380 | 149 | 56 | 22 |
| STAGE II-IV | 1434 | 1007 | 427 | 196 | 49 |
| METASTATIC | 380 | 279 | 101 | 84 | 35 |
| TOTAL | 8398 | 5900 | 2498 | 503 | 182 |

[A]The subset is obtained by fixing the categorical variables: propsol = 1, papnr = 0, Ascites = 0, Shadows = 0, loc10 = 0

**Table S3** Distribution of different classes in the CRASH dataset

| CLASS | ALL | TRAIN | TEST | IN FIGURES S5-S6 (TRAIN)[A] | IN FIGURES S7-8 (TEST)[A] |
|---|---|---|---|---|---|
| HEAD INJURY | 2309 | 1646 | 663 | 32 | 12 |
| NO DEATH | 10022 | 6998 | 3024 | 37 | 19 |
| OTHER | 217 | 139 | 78 | 3 | 3 |
| TOTAL | 12548 | 8783 | 3765 | 72 | 34 |

[A]The subset is obtained by fixing the categorical variables: sex=Female, gcsEyeOpening=1 None, pupilReact=One Reacts

**Table S4** Distribution of different classes in the IST dataset

| CLASS | ALL | TRAIN | TEST | IN FIGURES S10-S11 (TRAIN)[A] | IN FIGURES S12-13(TEST)[A] |
|---|---|---|---|---|---|
| HAEMORRHAGIC STROKE | 439 | 318 | 121 | 3 | 1 |
| INDETERMINATE STROKE | 736 | 530 | 206 | 18 | 8 |
| ISCHAEMIC STROKE | 13622 | 9507 | 4115 | 335 | 147 |
| NOT A STROKE | 344 | 243 | 101 | 24 | 12 |
| TOTAL | 15141 | 10598 | 4543 | 380 | 168 |

[A]The subset is obtained by fixing the categorical variables: RCONSC=F, RDEF1=N, RDEF2=N, RDEF3=N, RDEF4=N, RDEF5=N

**Table S5**     Coefficients and simulation factors

| Distribution | Predictors | Correlation | AUC | Strength | Coefficients |
|---|---|---|---|---|---|
| Continuous | 4 | 0 | 0.75 | Balanced | intercept -1.6; betas 0.51 |
| Continuous | 4 | 0 | 0.75 | Unbalanced | intercept -1.65; betas 0.97 and 0.24 |
| Continuous | 4 | 0.4 | 0.75 | Balanced | intercept -1.67; betas 0.35 |
| Continuous | 4 | 0.4 | 0.75 | Unbalanced | intercept -1.67; betas 0.74 and 0.18 |
| Continuous | 4 | 0 | 0.90 | Balanced | intercept -2.5; betas 1.2 |
| Continuous | 4 | 0 | 0.90 | Unbalanced | intercept -2.45; betas 2.2 and 0.55 |
| Continuous | 4 | 0.4 | 0.90 | Balanced | intercept -2.55; betas 0.85 |
| Continuous | 4 | 0.4 | 0.90 | Unbalanced | intercept -2.5; betas 1.71 and 0.43 |
| Binary | 4 | 0 | 0.75 | Balanced | intercept -3.9; betas 1.1 |
| Binary | 4 | 0 | 0.75 | Unbalanced | intercept -3.35; betas 1.91 and 0.48 |
| Binary | 4 | 0.4 | 0.75 | Balanced | intercept -3.28; betas 0.78 |
| Binary | 4 | 0.4 | 0.75 | Unbalanced | intercept -3.08; betas 1.59 and 0.40 |
| Binary | 4 | 0 | 0.90 | Balanced | intercept -8.0; betas 2.64 |
| Binary | 4 | 0 | 0.90 | Unbalanced | intercept -7.45; betas 5.02 and 1.26 |
| Binary | 4 | 0.4 | 0.90 | Balanced | intercept -6.4; betas 1.85 |
| Binary | 4 | 0.4 | 0.90 | Unbalanced | intercept -7.0; betas 4.37 and 1.09 |
| Continuous | 16 | 0 | 0.75 | Balanced | intercept -1.66; betas 0.26 |
| Continuous | 16 | 0 | 0.75 | Unbalanced | intercept -1.66; betas 0.48 and 0.12 |
| Continuous | 16 | 0.4 | 0.75 | Balanced | intercept -1.67; betas 0.10 |
| Continuous | 16 | 0.4 | 0.75 | Unbalanced | intercept -1.67; betas 0.22 and 0.054 |
| Continuous | 16 | 0 | 0.90 | Balanced | intercept -2.5; betas 0.61 |
| Continuous | 16 | 0 | 0.90 | Unbalanced | intercept -2.47; betas 1.09 and 0.27 |
| Continuous | 16 | 0.4 | 0.90 | Balanced | intercept -2.51; betas 0.23 |
| Continuous | 16 | 0.4 | 0.90 | Unbalanced | intercept -2.5; betas 0.50 and 0.126 |
| Binary | 16 | 0 | 0.75 | Balanced | intercept -5.8; betas 0.52 |
| Binary | 16 | 0 | 0.75 | Unbalanced | intercept -4.9; betas 0.93 and 0.23 |
| Binary | 16 | 0.4 | 0.75 | Balanced | intercept -3.5; betas 0.23 |
| Binary | 16 | 0.4 | 0.75 | Unbalanced | intercept -3.38; betas 0.50 and 0.124 |
| Binary | 16 | 0 | 0.90 | Balanced | intercept -12.6; betas 1.25 |
| Binary | 16 | 0 | 0.90 | Unbalanced | intercept -10.1; betas 2.17 and 0.54 |
| Binary | 16 | 0.4 | 0.90 | Balanced | intercept -7.0; betas 0.54 |
| Binary | 16 | 0.4 | 0.90 | Unbalanced | intercept -6.55 betas 1.14 and 0.28 |
| Continuous | 16 | 0 | 0.75 | Balanced | intercept -1.6; betas 0.51 and 0 |
| Continuous | 16 | 0 | 0.75 | Unbalanced | intercept -1.65; betas 0.97 , 0.24 and 0 |
| Continuous | 16 | 0.4 | 0.75 | Balanced | intercept -1.67; betas 0.35 and 0 |
| Continuous | 16 | 0.4 | 0.75 | Unbalanced | intercept -1.67; betas 0.74, 0.18, and 0 |
| Continuous | 16 | 0 | 0.90 | Balanced | intercept -2.5; betas 1.2 and 0 |
| Continuous | 16 | 0 | 0.90 | Unbalanced | intercept -2.45; betas 2.2, 0.55, and 0 |
| Continuous | 16 | 0.4 | 0.90 | Balanced | intercept -2.55; betas 0.85 and 0 |
| Continuous | 16 | 0.4 | 0.90 | Unbalanced | intercept -2.5; betas 1.71, 0.43, and 0 |
| Binary | 16 | 0 | 0.75 | Balanced | intercept -3.9; betas 1.1 and 0 |
| Binary | 16 | 0 | 0.75 | Unbalanced | intercept -3.35; betas 1.91, 0.48, and 0 |
| Binary | 16 | 0.4 | 0.75 | Balanced | intercept -3.28; betas 0.78 and 0 |
| Binary | 16 | 0.4 | 0.75 | Unbalanced | intercept -3.08; betas 1.59, 0.40, and 0 |

| Binary | 16 | 0 | 0.90 | Balanced | intercept -8.0; betas 2.64 and 0 |
| Binary | 16 | 0 | 0.90 | Unbalanced | intercept -7.45; betas 5.02, 1.26, and 0 |
| Binary | 16 | 0.4 | 0.90 | Balanced | intercept -6.4; betas 1.85 and 0 |
| Binary | 4 | 0.4 | 0.90 | Unbalanced | intercept -7.0; betas 4.37 and 1.09 |

(https://github.com/Goorbergh/resampling_techniquesCPM/blob/main/Simulation_study/Pre_simulation/Script/optim_beta.R)

**Table S6**     Main simulation results. Interquartile Range (IQR); Area under the roc curve (AUC).

| Scenario | Median Training AUC (IQR) | Median Test AUC (IQR) | Median Training Slope (IQR) | Median Test Slope (IQR) | Mean Variance (SD) | Mean Squared Bias (SD) | Mean Squared Error (SD) |
|---|---|---|---|---|---|---|---|
| b_16b_75_0_bal_20_200 | 1 (0) | 0.662 (0.024) | 4.592 (0.484) | 0.765 (0.128) | 0.011 (0.004) | 0.006 (0.01) | 0.017 (0.012) |
| b_16b_75_0_bal_20_4000 | 0.999 (0) | 0.703 (0.005) | 8.178 (0.8) | 0.857 (0.027) | 0.008 (0.004) | 0.002 (0.003) | 0.01 (0.006) |
| b_16b_75_0_bal_2_200 | 0.953 (0.016) | 0.682 (0.023) | 5.155 (1.167) | 1.284 (0.272) | 0.005 (0.001) | 0.009 (0.017) | 0.013 (0.018) |
| b_16b_75_0_bal_2_4000 | 0.921 (0.004) | 0.727 (0.004) | 3.409 (0.171) | 1.288 (0.06) | 0.002 (0.001) | 0.003 (0.006) | 0.006 (0.007) |
| b_16b_75_0_unb_20_200 | 1 (0) | 0.664 (0.028) | 4.631 (0.505) | 0.757 (0.118) | 0.011 (0.004) | 0.005 (0.007) | 0.016 (0.009) |
| b_16b_75_0_unb_20_4000 | 0.999 (0) | 0.704 (0.004) | 7.994 (0.758) | 0.779 (0.026) | 0.008 (0.004) | 0.001 (0.001) | 0.009 (0.004) |
| b_16b_75_0_unb_2_200 | 0.949 (0.017) | 0.687 (0.028) | 4.859 (1.134) | 1.26 (0.244) | 0.004 (0.002) | 0.007 (0.013) | 0.012 (0.014) |
| b_16b_75_0_unb_2_4000 | 0.903 (0.005) | 0.729 (0.003) | 2.647 (0.156) | 1.101 (0.058) | 0.002 (0.001) | 0.001 (0.003) | 0.004 (0.003) |
| b_16b_75_4_bal_20_200 | 0.994 (0.007) | 0.705 (0.022) | 3.919 (1.532) | 0.601 (0.112) | 0.012 (0.008) | 0.001 (0.001) | 0.013 (0.008) |
| b_16b_75_4_bal_20_4000 | 0.967 (0.005) | 0.722 (0.004) | 2.924 (0.244) | 0.733 (0.035) | 0.007 (0.005) | 0 (0) | 0.007 (0.005) |
| b_16b_75_4_bal_2_200 | 0.917 (0.018) | 0.73 (0.013) | 2.363 (0.548) | 0.893 (0.182) | 0.005 (0.004) | 0.001 (0.001) | 0.006 (0.004) |
| b_16b_75_4_bal_2_4000 | 0.882 (0.006) | 0.741 (0.002) | 1.897 (0.097) | 0.914 (0.054) | 0.003 (0.002) | 0 (0) | 0.003 (0.002) |
| b_16b_75_4_unb_20_200 | 0.995 (0.007) | 0.698 (0.022) | 3.946 (1.55) | 0.589 (0.115) | 0.013 (0.008) | 0.001 (0.001) | 0.014 (0.008) |
| b_16b_75_4_unb_20_4000 | 0.966 (0.005) | 0.718 (0.004) | 2.924 (0.234) | 0.718 (0.032) | 0.007 (0.005) | 0 (0) | 0.008 (0.005) |
| b_16b_75_4_unb_2_200 | 0.917 (0.018) | 0.723 (0.015) | 2.426 (0.593) | 0.884 (0.184) | 0.006 (0.004) | 0.001 (0.001) | 0.007 (0.004) |
| b_16b_75_4_unb_2_4000 | 0.878 (0.006) | 0.736 (0.002) | 1.895 (0.109) | 0.901 (0.053) | 0.003 (0.002) | 0 (0) | 0.003 (0.002) |
| b_16b_90_0_bal_20_200 | 1 (0) | 0.823 (0.021) | 4.101 (0.446) | 1.522 (0.219) | 0.01 (0.005) | 0.024 (0.038) | 0.033 (0.042) |
| b_16b_90_0_bal_20_4000 | 0.999 (0) | 0.873 (0.003) | 7.413 (0.802) | 1.46 (0.042) | 0.006 (0.005) | 0.01 (0.017) | 0.017 (0.02) |
| b_16b_90_0_bal_2_200 | 0.979 (0.011) | 0.833 (0.022) | 5.602 (1.368) | 2.337 (0.459) | 0.005 (0.002) | 0.034 (0.059) | 0.038 (0.061) |
| b_16b_90_0_bal_2_4000 | 0.968 (0.003) | 0.885 (0.003) | 3.714 (0.179) | 2.054 (0.08) | 0.003 (0.002) | 0.016 (0.029) | 0.019 (0.03) |
| b_16b_90_0_unb_20_200 | 1 (0) | 0.839 (0.017) | 4.251 (0.506) | 1.441 (0.185) | 0.009 (0.006) | 0.016 (0.026) | 0.026 (0.029) |
| b_16b_90_0_unb_20_4000 | 0.998 (0) | 0.877 (0.002) | 6.251 (0.577) | 1.145 (0.039) | 0.006 (0.005) | 0.004 (0.006) | 0.01 (0.01) |
| b_16b_90_0_unb_2_200 | 0.973 (0.012) | 0.854 (0.017) | 4.451 (1.047) | 2.179 (0.362) | 0.004 (0.002) | 0.024 (0.042) | 0.029 (0.044) |
| b_16b_90_0_unb_2_4000 | 0.951 (0.004) | 0.887 (0.003) | 2.217 (0.078) | 1.432 (0.058) | 0.002 (0.002) | 0.006 (0.011) | 0.008 (0.012) |
| b_16b_90_4_bal_20_200 | 0.995 (0.006) | 0.879 (0.01) | 3.219 (1.456) | 0.976 (0.139) | 0.01 (0.01) | 0.002 (0.002) | 0.012 (0.012) |
| b_16b_90_4_bal_20_4000 | 0.978 (0.002) | 0.89 (0.002) | 2.331 (0.103) | 0.968 (0.04) | 0.006 (0.006) | 0.001 (0.001) | 0.006 (0.006) |
| b_16b_90_4_bal_2_200 | 0.963 (0.016) | 0.887 (0.007) | 2.16 (0.315) | 1.275 (0.197) | 0.005 (0.005) | 0.005 (0.006) | 0.01 (0.011) |
| b_16b_90_4_bal_2_4000 | 0.947 (0.004) | 0.897 (0.001) | 1.601 (0.035) | 1.101 (0.047) | 0.002 (0.003) | 0.001 (0.001) | 0.003 (0.004) |
| b_16b_90_4_unb_20_200 | 0.995 (0.005) | 0.869 (0.012) | 3.164 (1.113) | 0.973 (0.149) | 0.01 (0.01) | 0.003 (0.003) | 0.013 (0.012) |
| b_16b_90_4_unb_20_4000 | 0.977 (0.002) | 0.882 (0.002) | 2.266 (0.111) | 0.929 (0.038) | 0.006 (0.006) | 0 (0) | 0.006 (0.006) |
| b_16b_90_4_unb_2_200 | 0.958 (0.016) | 0.88 (0.009) | 2.114 (0.294) | 1.267 (0.203) | 0.005 (0.005) | 0.005 (0.007) | 0.01 (0.011) |
| b_16b_90_4_unb_2_4000 | 0.941 (0.004) | 0.89 (0.001) | 1.524 (0.034) | 1.055 (0.046) | 0.002 (0.003) | 0.001 (0.001) | 0.003 (0.003) |
| b_16c_75_0_bal_20_200 | 1 (0) | 0.637 (0.025) | 3.385 (0.174) | 0.668 (0.187) | 0.008 (0.003) | 0.011 (0.016) | 0.019 (0.018) |
| b_16c_75_0_bal_20_4000 | 1 (0) | 0.701 (0.005) | 5.639 (0.071) | 1.148 (0.071) | 0.002 (0.001) | 0.006 (0.01) | 0.009 (0.01) |
| b_16c_75_0_bal_2_200 | 1 (0.001) | 0.635 (0.025) | 9.274 (3.05) | 0.708 (0.214) | 0.007 (0.002) | 0.012 (0.018) | 0.019 (0.019) |
| b_16c_75_0_bal_2_4000 | 1 (0) | 0.702 (0.006) | 19.411 (3.012) | 1.177 (0.076) | 0.002 (0.001) | 0.007 (0.01) | 0.009 (0.01) |
| b_16c_75_0_unb_20_200 | 1 (0) | 0.645 (0.029) | 3.384 (0.184) | 0.701 (0.192) | 0.008 (0.003) | 0.01 (0.015) | 0.018 (0.016) |
| b_16c_75_0_unb_20_4000 | 1 (0) | 0.71 (0.004) | 5.89 (0.103) | 1.034 (0.055) | 0.003 (0.001) | 0.005 (0.007) | 0.007 (0.008) |
| b_16c_75_0_unb_2_200 | 0.999 (0.001) | 0.644 (0.03) | 8.986 (3.157) | 0.747 (0.22) | 0.007 (0.002) | 0.011 (0.016) | 0.018 (0.017) |
| b_16c_75_0_unb_2_4000 | 1 (0) | 0.712 (0.004) | 14.801 (2.427) | 1.055 (0.058) | 0.002 (0.001) | 0.005 (0.007) | 0.007 (0.008) |
| b_16c_75_4_bal_20_200 | 1 (0) | 0.713 (0.018) | 3.583 (0.216) | 0.738 (0.123) | 0.007 (0.004) | 0.002 (0.003) | 0.009 (0.005) |
| b_16c_75_4_bal_20_4000 | 1 (0) | 0.734 (0.003) | 6.563 (0.186) | 0.9 (0.049) | 0.002 (0.001) | 0.001 (0.003) | 0.004 (0.003) |
| b_16c_75_4_bal_2_200 | 0.992 (0.005) | 0.719 (0.017) | 5.052 (1.832) | 0.788 (0.13) | 0.006 (0.003) | 0.002 (0.003) | 0.008 (0.004) |
| b_16c_75_4_bal_2_4000 | 0.998 (0.001) | 0.735 (0.003) | 8.486 (1.065) | 0.91 (0.049) | 0.002 (0.001) | 0.002 (0.003) | 0.003 (0.003) |
| b_16c_75_4_unb_20_200 | 1 (0.019) | 0.706 (0.019) | 3.57 (0.22) | 0.724 (0.122) | 0.007 (0.004) | 0.002 (0.004) | 0.01 (0.006) |
| b_16c_75_4_unb_20_4000 | 1 (0) | 0.727 (0.003) | 6.503 (0.181) | 0.887 (0.049) | 0.002 (0.001) | 0.002 (0.003) | 0.004 (0.003) |
| b_16c_75_4_unb_2_200 | 0.993 (0.005) | 0.71 (0.018) | 5.126 (1.924) | 0.773 (0.136) | 0.006 (0.003) | 0.002 (0.004) | 0.008 (0.005) |
| b_16c_75_4_unb_2_4000 | 0.998 (0.001) | 0.729 (0.003) | 8.78 (1.158) | 0.897 (0.048) | 0.002 (0.001) | 0.002 (0.003) | 0.004 (0.003) |
| b_16c_90_0_bal_20_200 | 1 (0) | 0.781 (0.026) | 3.213 (0.167) | 1.458 (0.341) | 0.008 (0.003) | 0.035 (0.058) | 0.043 (0.06) |
| b_16c_90_0_bal_20_4000 | 1 (0) | 0.858 (0.005) | 5.396 (0.074) | 2.059 (0.104) | 0.002 (0.001) | 0.023 (0.038) | 0.026 (0.039) |
| b_16c_90_0_bal_2_200 | 1 (0.001) | 0.773 (0.028) | 8.515 (2.377) | 1.526 (0.397) | 0.007 (0.003) | 0.038 (0.063) | 0.045 (0.064) |
| b_16c_90_0_bal_2_4000 | 1 (0) | 0.857 (0.005) | 14.996 (2.26) | 2.152 (0.116) | 0.002 (0.001) | 0.025 (0.041) | 0.027 (0.041) |
| b_16c_90_0_unb_20_200 | 1 (0) | 0.8 (0.027) | 3.197 (0.163) | 1.47 (0.296) | 0.008 (0.004) | 0.028 (0.046) | 0.036 (0.048) |
| b_16c_90_0_unb_20_4000 | 1 (0) | 0.866 (0.003) | 5.983 (0.136) | 1.54 (0.072) | 0.003 (0.002) | 0.013 (0.022) | 0.016 (0.023) |
| b_16c_90_0_unb_2_200 | 1 (0.001) | 0.798 (0.029) | 8.071 (2.626) | 1.569 (0.354) | 0.007 (0.003) | 0.031 (0.05) | 0.038 (0.051) |
| b_16c_90_0_unb_2_4000 | 0.999 (0) | 0.866 (0.003) | 8.925 (1.088) | 1.588 (0.077) | 0.002 (0.002) | 0.014 (0.023) | 0.016 (0.024) |
| b_16c_90_4_bal_20_200 | 1 (0) | 0.88 (0.007) | 3.3 (0.252) | 1.109 (0.146) | 0.006 (0.005) | 0.004 (0.006) | 0.009 (0.009) |
| b_16c_90_4_bal_20_4000 | 1 (0) | 0.891 (0.001) | 6.976 (0.767) | 1.098 (0.045) | 0.002 (0.002) | 0.002 (0.004) | 0.004 (0.005) |
| b_16c_90_4_bal_2_200 | 0.995 (0.003) | 0.882 (0.006) | 4.007 (0.974) | 1.209 (0.139) | 0.005 (0.004) | 0.005 (0.008) | 0.01 (0.01) |
| b_16c_90_4_bal_2_4000 | 0.996 (0) | 0.892 (0.001) | 4.655 (0.308) | 1.121 (0.041) | 0.001 (0.001) | 0.002 (0.004) | 0.003 (0.005) |
| b_16c_90_4_unb_20_200 | 1 (0) | 0.873 (0.008) | 3.316 (0.264) | 1.094 (0.15) | 0.006 (0.005) | 0.005 (0.009) | 0.01 (0.011) |
| b_16c_90_4_unb_20_4000 | 1 (0) | 0.887 (0.001) | 7.056 (0.803) | 1.081 (0.048) | 0.002 (0.002) | 0.002 (0.005) | 0.004 (0.005) |
| b_16c_90_4_unb_2_200 | 0.995 (0.003) | 0.875 (0.008) | 4.014 (1.04) | 1.188 (0.142) | 0.005 (0.004) | 0.006 (0.01) | 0.01 (0.012) |
| b_16c_90_4_unb_2_4000 | 0.996 (0) | 0.887 (0.001) | 4.64 (0.32) | 1.102 (0.044) | 0.001 (0.001) | 0.002 (0.005) | 0.004 (0.005) |
| b_4b_75_0_bal_20_200 | 0.803 (0.047) | 0.736 (0.022) | 1.321 (0.083) | 0.884 (0.185) | 0.006 (0.003) | 0.001 (0.002) | 0.007 (0.004) |
| b_4b_75_0_bal_20_4000 | 0.762 (0.013) | 0.756 (0.001) | 1.283 (0.023) | 1.259 (0.072) | 0.001 (0) | 0.001 (0.002) | 0.002 (0.002) |
| b_4b_75_0_bal_2_200 | 0.801 (0.047) | 0.738 (0.02) | 1.351 (0.097) | 0.939 (0.201) | 0.006 (0.003) | 0.001 (0.002) | 0.007 (0.004) |
| b_4b_75_0_bal_2_4000 | 0.762 (0.013) | 0.756 (0.001) | 1.283 (0.023) | 1.259 (0.072) | 0.001 (0) | 0.001 (0.002) | 0.002 (0.002) |
| b_4b_75_0_bal_noise_20_200 | 1 (0) | 0.684 (0.026) | 4.652 (0.512) | 0.816 (0.119) | 0.011 (0.004) | 0.005 (0.007) | 0.016 (0.009) |

| Scenario | Median Training AUC (IQR) | Median Test AUC (IQR) | Median Training Slope (IQR) | Median Test Slope (IQR) | Mean Variance (SD) | Mean Squared Bias (SD) | Mean Squared Error (SD) |
|---|---|---|---|---|---|---|---|
| b_4b_75_0_bal_noise_20_4000 | 0.998 (0) | 0.724 (0.004) | 7.715 (0.708) | 0.797 (0.027) | 0.008 (0.004) | 0.001 (0.001) | 0.009 (0.004) |
| b_4b_75_0_bal_noise_2_200 | 0.948 (0.018) | 0.709 (0.025) | 4.497 (1.119) | 1.332 (0.234) | 0.004 (0.002) | 0.007 (0.013) | 0.012 (0.014) |
| b_4b_75_0_bal_noise_2_4000 | 0.903 (0.005) | 0.746 (0.002) | 2.368 (0.123) | 1.081 (0.056) | 0.002 (0.001) | 0.001 (0.002) | 0.003 (0.002) |
| b_4b_75_0_unb_20_200 | 0.799 (0.046) | 0.729 (0.019) | 1.27 (0.084) | 0.819 (0.172) | 0.006 (0.003) | 0 (0) | 0.006 (0.004) |
| b_4b_75_0_unb_20_4000 | 0.755 (0.011) | 0.752 (0.001) | 1.157 (0.018) | 1.136 (0.066) | 0 (0) | 0 (0.001) | 0.001 (0.001) |
| b_4b_75_0_unb_2_200 | 0.796 (0.046) | 0.731 (0.018) | 1.29 (0.089) | 0.872 (0.185) | 0.005 (0.003) | 0 (0.001) | 0.006 (0.003) |
| b_4b_75_0_unb_2_4000 | 0.755 (0.011) | 0.752 (0.001) | 1.157 (0.018) | 1.136 (0.066) | 0 (0) | 0 (0.001) | 0.001 (0.001) |
| b_4b_75_0_unb_noise_20_200 | 1 (0) | 0.688 (0.029) | 4.688 (0.505) | 0.789 (0.105) | 0.01 (0.005) | 0.003 (0.003) | 0.013 (0.005) |
| b_4b_75_0_unb_noise_20_4000 | 0.999 (0) | 0.719 (0.004) | 8.17 (0.744) | 0.776 (0.026) | 0.008 (0.004) | 0 (0) | 0.008 (0.004) |
| b_4b_75_0_unb_noise_2_200 | 0.94 (0.018) | 0.713 (0.022) | 4.241 (1.083) | 1.262 (0.226) | 0.004 (0.002) | 0.005 (0.005) | 0.009 (0.006) |
| b_4b_75_0_unb_noise_2_4000 | 0.897 (0.005) | 0.739 (0.003) | 2.503 (0.157) | 1.046 (0.055) | 0.002 (0.001) | 0.001 (0.001) | 0.003 (0.002) |
| b_4b_75_4_bal_20_200 | 0.8 (0.048) | 0.73 (0.02) | 1.211 (0.08) | 0.765 (0.195) | 0.006 (0.003) | 0 (0) | 0.006 (0.003) |
| b_4b_75_4_bal_20_4000 | 0.757 (0.013) | 0.752 (0) | 1.134 (0.012) | 1.108 (0.07) | 0 (0) | 0 (0) | 0.001 (0) |
| b_4b_75_4_bal_2_200 | 0.794 (0.048) | 0.736 (0.018) | 1.225 (0.085) | 0.843 (0.215) | 0.005 (0.003) | 0 (0) | 0.005 (0.003) |
| b_4b_75_4_bal_2_4000 | 0.757 (0.013) | 0.752 (0) | 1.134 (0.012) | 1.108 (0.07) | 0 (0) | 0 (0) | 0.001 (0) |
| b_4b_75_4_bal_noise_20_200 | 0.995 (0.007) | 0.695 (0.025) | 4.05 (1.433) | 0.589 (0.121) | 0.012 (0.008) | 0.002 (0.002) | 0.014 (0.008) |
| b_4b_75_4_bal_noise_20_4000 | 0.969 (0.004) | 0.721 (0.004) | 2.977 (0.22) | 0.722 (0.032) | 0.007 (0.005) | 0 (0) | 0.008 (0.005) |
| b_4b_75_4_bal_noise_2_200 | 0.923 (0.019) | 0.72 (0.017) | 2.671 (0.87) | 0.894 (0.218) | 0.005 (0.004) | 0.002 (0.004) | 0.008 (0.006) |
| b_4b_75_4_bal_noise_2_4000 | 0.88 (0.006) | 0.739 (0.002) | 1.932 (0.108) | 0.911 (0.053) | 0.003 (0.002) | 0 (0.001) | 0.003 (0.002) |
| b_4b_75_4_unb_20_200 | 0.796 (0.046) | 0.724 (0.02) | 1.209 (0.088) | 0.775 (0.191) | 0.006 (0.004) | 0 (0) | 0.006 (0.004) |
| b_4b_75_4_unb_20_4000 | 0.752 (0.012) | 0.749 (0.001) | 1.102 (0.011) | 1.079 (0.068) | 0 (0) | 0 (0) | 0 (0) |
| b_4b_75_4_unb_2_200 | 0.79 (0.047) | 0.731 (0.018) | 1.223 (0.094) | 0.845 (0.219) | 0.005 (0.003) | 0 (0) | 0.005 (0.003) |
| b_4b_75_4_unb_2_4000 | 0.752 (0.012) | 0.749 (0.001) | 1.102 (0.011) | 1.079 (0.068) | 0 (0) | 0 (0) | 0 (0) |
| b_4b_75_4_unb_noise_20_200 | 0.996 (0.007) | 0.691 (0.027) | 4.06 (1.407) | 0.591 (0.138) | 0.013 (0.008) | 0.002 (0.002) | 0.015 (0.007) |
| b_4b_75_4_unb_noise_20_4000 | 0.969 (0.004) | 0.718 (0.004) | 3.137 (0.216) | 0.726 (0.031) | 0.007 (0.005) | 0 (0) | 0.008 (0.005) |
| b_4b_75_4_unb_noise_2_200 | 0.922 (0.018) | 0.718 (0.02) | 2.891 (0.968) | 0.924 (0.225) | 0.005 (0.004) | 0.002 (0.003) | 0.008 (0.004) |
| b_4b_75_4_unb_noise_2_4000 | 0.877 (0.006) | 0.735 (0.003) | 2.027 (0.112) | 0.917 (0.053) | 0.003 (0.002) | 0 (0) | 0.003 (0.002) |
| b_4b_90_0_bal_20_200 | 0.925 (0.027) | 0.899 (0.002) | 1.576 (0.167) | 1.331 (0.198) | 0.005 (0.005) | 0.004 (0.005) | 0.009 (0.008) |
| b_4b_90_0_bal_20_4000 | 0.903 (0.007) | 0.899 (0.001) | 1.601 (0.049) | 1.587 (0.063) | 0.001 (0.001) | 0.004 (0.006) | 0.005 (0.006) |
| b_4b_90_0_bal_2_200 | 0.924 (0.027) | 0.899 (0.002) | 1.645 (0.193) | 1.412 (0.22) | 0.005 (0.005) | 0.005 (0.006) | 0.01 (0.009) |
| b_4b_90_0_bal_2_4000 | 0.903 (0.007) | 0.899 (0.001) | 1.601 (0.049) | 1.587 (0.063) | 0.001 (0.001) | 0.004 (0.006) | 0.005 (0.006) |
| b_4b_90_0_bal_noise_20_200 | 1 (0) | 0.867 (0.016) | 4.265 (0.558) | 1.514 (0.166) | 0.009 (0.006) | 0.014 (0.02) | 0.023 (0.023) |
| b_4b_90_0_bal_noise_20_4000 | 0.999 (0) | 0.894 (0.001) | 5.954 (0.53) | 1.137 (0.032) | 0.006 (0.006) | 0.002 (0.002) | 0.007 (0.006) |
| b_4b_90_0_bal_noise_2_200 | 0.975 (0.01) | 0.883 (0.016) | 4.266 (1.005) | 2.3 (0.341) | 0.004 (0.002) | 0.022 (0.035) | 0.026 (0.036) |
| b_4b_90_0_bal_noise_2_4000 | 0.965 (0.004) | 0.898 (0.001) | 2.099 (0.067) | 1.406 (0.044) | 0.002 (0.002) | 0.003 (0.004) | 0.005 (0.005) |
| b_4b_90_0_unb_20_200 | 0.912 (0.027) | 0.892 (0.008) | 1.42 (0.124) | 1.188 (0.194) | 0.005 (0.005) | 0.002 (0.003) | 0.006 (0.006) |
| b_4b_90_0_unb_20_4000 | 0.9 (0.007) | 0.895 (0) | 1.428 (0.044) | 1.391 (0.066) | 0 (0) | 0.002 (0.003) | 0.002 (0.003) |
| b_4b_90_0_unb_2_200 | 0.911 (0.027) | 0.894 (0.007) | 1.465 (0.144) | 1.252 (0.2) | 0.004 (0.004) | 0.002 (0.004) | 0.006 (0.006) |
| b_4b_90_0_unb_2_4000 | 0.9 (0.007) | 0.895 (0) | 1.428 (0.044) | 1.391 (0.066) | 0 (0) | 0.002 (0.003) | 0.002 (0.003) |
| b_4b_90_0_unb_noise_20_200 | 1 (0) | 0.864 (0.013) | 4.396 (0.519) | 1.384 (0.134) | 0.008 (0.007) | 0.009 (0.014) | 0.018 (0.019) |
| b_4b_90_0_unb_noise_20_4000 | 0.999 (0) | 0.886 (0.002) | 6.637 (0.654) | 1.072 (0.03) | 0.006 (0.006) | 0.001 (0.002) | 0.007 (0.007) |
| b_4b_90_0_unb_noise_2_200 | 0.971 (0.012) | 0.877 (0.013) | 4.388 (1.159) | 2.064 (0.243) | 0.004 (0.003) | 0.015 (0.025) | 0.019 (0.027) |
| b_4b_90_0_unb_noise_2_4000 | 0.95 (0.003) | 0.895 (0.001) | 2.141 (0.095) | 1.328 (0.047) | 0.002 (0.002) | 0.002 (0.003) | 0.004 (0.004) |
| b_4b_90_4_bal_20_200 | 0.918 (0.027) | 0.897 (0.007) | 1.308 (0.091) | 1.088 (0.187) | 0.005 (0.005) | 0.001 (0.002) | 0.006 (0.005) |
| b_4b_90_4_bal_20_4000 | 0.903 (0.007) | 0.9 (0) | 1.29 (0.028) | 1.272 (0.054) | 0.001 (0.001) | 0.001 (0.002) | 0.002 (0.002) |
| b_4b_90_4_bal_2_200 | 0.916 (0.028) | 0.896 (0.007) | 1.323 (0.095) | 1.148 (0.206) | 0.004 (0.004) | 0.002 (0.002) | 0.006 (0.004) |
| b_4b_90_4_bal_2_4000 | 0.903 (0.007) | 0.9 (0) | 1.29 (0.028) | 1.272 (0.054) | 0.001 (0.001) | 0.001 (0.002) | 0.002 (0.002) |
| b_4b_90_4_bal_noise_20_200 | 0.995 (0.004) | 0.876 (0.012) | 3.237 (1.22) | 1.059 (0.179) | 0.01 (0.009) | 0.006 (0.011) | 0.016 (0.017) |
| b_4b_90_4_bal_noise_20_4000 | 0.979 (0.002) | 0.893 (0.001) | 2.136 (0.1) | 0.972 (0.04) | 0.005 (0.005) | 0.001 (0.002) | 0.006 (0.006) |
| b_4b_90_4_bal_noise_2_200 | 0.959 (0.015) | 0.888 (0.01) | 2.12 (0.291) | 1.367 (0.255) | 0.005 (0.004) | 0.009 (0.019) | 0.013 (0.021) |
| b_4b_90_4_bal_noise_2_4000 | 0.948 (0.004) | 0.899 (0.001) | 1.516 (0.033) | 1.103 (0.046) | 0.002 (0.002) | 0.001 (0.003) | 0.003 (0.004) |
| b_4b_90_4_unb_20_200 | 0.909 (0.027) | 0.894 (0.007) | 1.306 (0.086) | 1.101 (0.193) | 0.004 (0.005) | 0.001 (0.001) | 0.005 (0.005) |
| b_4b_90_4_unb_20_4000 | 0.9 (0.007) | 0.898 (0) | 1.295 (0.03) | 1.273 (0.064) | 0 (0) | 0.001 (0.001) | 0.001 (0.001) |
| b_4b_90_4_unb_2_200 | 0.908 (0.027) | 0.895 (0.005) | 1.33 (0.091) | 1.173 (0.221) | 0.004 (0.004) | 0.001 (0.001) | 0.005 (0.004) |
| b_4b_90_4_unb_2_4000 | 0.9 (0.007) | 0.898 (0) | 1.295 (0.03) | 1.273 (0.064) | 0 (0) | 0.001 (0.001) | 0.001 (0.001) |
| b_4b_90_4_unb_noise_20_200 | 0.996 (0.004) | 0.873 (0.011) | 3.719 (1.539) | 1.068 (0.173) | 0.01 (0.01) | 0.004 (0.008) | 0.014 (0.014) |
| b_4b_90_4_unb_noise_20_4000 | 0.98 (0.002) | 0.889 (0.002) | 2.649 (0.195) | 0.97 (0.04) | 0.005 (0.006) | 0.001 (0.001) | 0.006 (0.006) |
| b_4b_90_4_unb_noise_2_200 | 0.956 (0.013) | 0.884 (0.009) | 2.265 (0.425) | 1.387 (0.286) | 0.004 (0.005) | 0.007 (0.013) | 0.012 (0.015) |
| b_4b_90_4_unb_noise_2_4000 | 0.94 (0.004) | 0.895 (0.001) | 1.675 (0.062) | 1.121 (0.052) | 0.002 (0.002) | 0.001 (0.002) | 0.003 (0.003) |
| b_4c_75_0_bal_20_200 | 1 (0) | 0.672 (0.023) | 3.984 (0.304) | 0.483 (0.107) | 0.017 (0.009) | 0.002 (0.005) | 0.02 (0.011) |
| b_4c_75_0_bal_20_4000 | 1 (0) | 0.705 (0.005) | 7.598 (0.267) | 0.656 (0.031) | 0.01 (0.006) | 0.001 (0.002) | 0.01 (0.006) |
| b_4c_75_0_bal_2_200 | 0.981 (0.008) | 0.685 (0.023) | 4.096 (1.269) | 0.621 (0.136) | 0.011 (0.006) | 0.003 (0.006) | 0.015 (0.01) |
| b_4c_75_0_bal_2_4000 | 0.99 (0.002) | 0.716 (0.004) | 5.81 (0.585) | 0.76 (0.038) | 0.006 (0.004) | 0.001 (0.003) | 0.007 (0.005) |
| b_4c_75_0_bal_noise_20_200 | 1 (0) | 0.643 (0.028) | 3.423 (0.176) | 0.695 (0.189) | 0.008 (0.003) | 0.009 (0.014) | 0.018 (0.015) |
| b_4c_75_0_bal_noise_20_4000 | 1 (0) | 0.705 (0.004) | 6.044 (0.104) | 0.96 (0.048) | 0.003 (0.002) | 0.004 (0.006) | 0.007 (0.007) |
| b_4c_75_0_bal_noise_2_200 | 0.999 (0.001) | 0.643 (0.029) | 8.81 (3.01) | 0.742 (0.209) | 0.007 (0.002) | 0.01 (0.015) | 0.017 (0.015) |
| b_4c_75_0_bal_noise_2_4000 | 1 (0) | 0.707 (0.004) | 13.088 (2.055) | 0.98 (0.049) | 0.003 (0.001) | 0.004 (0.006) | 0.006 (0.007) |
| b_4c_75_0_unb_20_200 | 1 (0) | 0.684 (0.025) | 4.013 (0.328) | 0.497 (0.099) | 0.017 (0.01) | 0.002 (0.005) | 0.019 (0.011) |
| b_4c_75_0_unb_20_4000 | 1 (0) | 0.715 (0.004) | 7.703 (0.294) | 0.657 (0.034) | 0.009 (0.006) | 0.001 (0.002) | 0.01 (0.006) |
| b_4c_75_0_unb_2_200 | 0.979 (0.009) | 0.697 (0.024) | 3.714 (1.258) | 0.632 (0.128) | 0.011 (0.006) | 0.003 (0.006) | 0.014 (0.009) |
| b_4c_75_0_unb_2_4000 | 0.99 (0.002) | 0.726 (0.004) | 5.611 (0.571) | 0.756 (0.04) | 0.006 (0.004) | 0.001 (0.003) | 0.007 (0.005) |

| Scenario | Median Training AUC (IQR) | Median Test AUC (IQR) | Median Training Slope (IQR) | Median Test Slope (IQR) | Mean Variance (SD) | Mean Squared Bias (SD) | Mean Squared Error (SD) |
|---|---|---|---|---|---|---|---|
| b_4c_75_0_unb_noise_20_200 | 1 (0) | 0.664 (0.029) | 3.424 (0.189) | 0.748 (0.153) | 0.008 (0.003) | 0.007 (0.011) | 0.015 (0.013) |
| b_4c_75_0_unb_noise_20_4000 | 1 (0) | 0.72 (0.004) | 6.23 (0.155) | 0.947 (0.05) | 0.003 (0.002) | 0.003 (0.005) | 0.006 (0.006) |
| b_4c_75_0_unb_noise_2_200 | 0.999 (0.002) | 0.665 (0.03) | 7.852 (2.957) | 0.802 (0.176) | 0.007 (0.003) | 0.008 (0.012) | 0.015 (0.013) |
| b_4c_75_0_unb_noise_2_4000 | 0.999 (0) | 0.721 (0.005) | 11.246 (1.952) | 0.962 (0.05) | 0.002 (0.001) | 0.003 (0.006) | 0.005 (0.006) |
| b_4c_75_4_bal_20_200 | 1 (0) | 0.69 (0.024) | 4.035 (0.349) | 0.452 (0.097) | 0.016 (0.009) | 0.001 (0.003) | 0.018 (0.01) |
| b_4c_75_4_bal_20_4000 | 1 (0) | 0.712 (0.005) | 7.667 (0.315) | 0.636 (0.039) | 0.009 (0.006) | 0 (0.002) | 0.01 (0.006) |
| b_4c_75_4_bal_2_200 | 0.978 (0.008) | 0.704 (0.021) | 3.483 (1.206) | 0.595 (0.122) | 0.01 (0.006) | 0.002 (0.003) | 0.012 (0.007) |
| b_4c_75_4_bal_2_4000 | 0.991 (0.002) | 0.722 (0.004) | 5.712 (0.6) | 0.739 (0.042) | 0.006 (0.004) | 0.001 (0.002) | 0.006 (0.004) |
| b_4c_75_4_bal_noise_20_200 | 1 (0) | 0.698 (0.021) | 3.552 (0.213) | 0.727 (0.132) | 0.008 (0.004) | 0.004 (0.006) | 0.011 (0.008) |
| b_4c_75_4_bal_noise_20_4000 | 1 (0) | 0.726 (0.003) | 6.42 (0.172) | 0.909 (0.051) | 0.002 (0.001) | 0.002 (0.004) | 0.005 (0.004) |
| b_4c_75_4_bal_noise_2_200 | 0.994 (0.005) | 0.702 (0.021) | 5.544 (2.123) | 0.774 (0.146) | 0.006 (0.003) | 0.004 (0.006) | 0.01 (0.008) |
| b_4c_75_4_bal_noise_2_4000 | 0.999 (0.001) | 0.728 (0.003) | 9.186 (1.259) | 0.92 (0.051) | 0.002 (0.001) | 0.002 (0.004) | 0.004 (0.004) |
| b_4c_75_4_unb_20_200 | 1 (0) | 0.685 (0.026) | 4.018 (0.329) | 0.454 (0.103) | 0.017 (0.01) | 0.001 (0.003) | 0.018 (0.01) |
| b_4c_75_4_unb_20_4000 | 1 (0) | 0.71 (0.005) | 7.671 (0.325) | 0.632 (0.037) | 0.009 (0.006) | 0 (0.002) | 0.01 (0.006) |
| b_4c_75_4_unb_2_200 | 0.978 (0.009) | 0.7 (0.024) | 3.557 (1.25) | 0.592 (0.124) | 0.011 (0.006) | 0.002 (0.004) | 0.012 (0.008) |
| b_4c_75_4_unb_2_4000 | 0.991 (0.002) | 0.72 (0.004) | 5.788 (0.598) | 0.736 (0.041) | 0.006 (0.004) | 0.001 (0.002) | 0.006 (0.005) |
| b_4c_75_4_unb_noise_20_200 | 1 (0) | 0.689 (0.024) | 3.528 (0.218) | 0.712 (0.141) | 0.008 (0.004) | 0.004 (0.007) | 0.012 (0.009) |
| b_4c_75_4_unb_noise_20_4000 | 1 (0) | 0.723 (0.004) | 6.37 (0.177) | 0.905 (0.051) | 0.002 (0.001) | 0.002 (0.004) | 0.005 (0.004) |
| b_4c_75_4_unb_noise_2_200 | 0.994 (0.005) | 0.693 (0.023) | 5.691 (2.316) | 0.76 (0.158) | 0.006 (0.003) | 0.005 (0.008) | 0.011 (0.009) |
| b_4c_75_4_unb_noise_2_4000 | 0.999 (0.001) | 0.725 (0.004) | 9.592 (1.482) | 0.917 (0.051) | 0.002 (0.001) | 0.002 (0.004) | 0.004 (0.004) |
| b_4c_90_0_bal_20_200 | 1 (0) | 0.842 (0.016) | 3.655 (0.338) | 0.873 (0.164) | 0.014 (0.012) | 0.008 (0.016) | 0.023 (0.024) |
| b_4c_90_0_bal_20_4000 | 1 (0) | 0.875 (0.003) | 7.046 (1.565) | 0.886 (0.039) | 0.008 (0.008) | 0.002 (0.006) | 0.009 (0.011) |
| b_4c_90_0_bal_2_200 | 0.989 (0.006) | 0.846 (0.017) | 3.68 (0.926) | 1.076 (0.223) | 0.01 (0.008) | 0.012 (0.021) | 0.022 (0.025) |
| b_4c_90_0_bal_2_4000 | 0.99 (0.001) | 0.881 (0.002) | 3.418 (0.201) | 0.986 (0.041) | 0.005 (0.005) | 0.002 (0.007) | 0.007 (0.009) |
| b_4c_90_0_bal_noise_20_200 | 1 (0) | 0.816 (0.025) | 3.184 (0.18) | 1.499 (0.321) | 0.007 (0.004) | 0.025 (0.041) | 0.033 (0.043) |
| b_4c_90_0_bal_noise_20_4000 | 1 (0) | 0.875 (0.003) | 6.304 (0.156) | 1.412 (0.057) | 0.003 (0.002) | 0.008 (0.015) | 0.011 (0.016) |
| b_4c_90_0_bal_noise_2_200 | 0.999 (0.001) | 0.814 (0.027) | 7.536 (2.514) | 1.608 (0.388) | 0.006 (0.003) | 0.028 (0.044) | 0.034 (0.046) |
| b_4c_90_0_bal_noise_2_4000 | 0.999 (0) | 0.876 (0.003) | 7.056 (0.642) | 1.454 (0.058) | 0.002 (0.002) | 0.009 (0.016) | 0.011 (0.017) |
| b_4c_90_0_unb_20_200 | 1 (0) | 0.855 (0.014) | 3.844 (0.418) | 0.807 (0.142) | 0.013 (0.012) | 0.004 (0.009) | 0.017 (0.017) |
| b_4c_90_0_unb_20_4000 | 1 (0) | 0.879 (0.002) | 6.095 (4.126) | 0.832 (0.039) | 0.007 (0.007) | 0.001 (0.003) | 0.008 (0.008) |
| b_4c_90_0_unb_2_200 | 0.985 (0.006) | 0.861 (0.014) | 2.651 (0.69) | 0.968 (0.182) | 0.009 (0.008) | 0.006 (0.011) | 0.015 (0.016) |
| b_4c_90_0_unb_2_4000 | 0.989 (0.001) | 0.885 (0.002) | 3.228 (0.221) | 0.918 (0.04) | 0.005 (0.005) | 0.001 (0.003) | 0.006 (0.006) |
| b_4c_90_0_unb_noise_20_200 | 1 (0) | 0.837 (0.02) | 3.256 (0.196) | 1.369 (0.248) | 0.007 (0.004) | 0.016 (0.025) | 0.023 (0.027) |
| b_4c_90_0_unb_noise_20_4000 | 1 (0) | 0.879 (0.002) | 6.996 (0.212) | 1.208 (0.054) | 0.002 (0.002) | 0.004 (0.009) | 0.007 (0.009) |
| b_4c_90_0_unb_noise_2_200 | 0.997 (0.003) | 0.836 (0.021) | 5.39 (2.142) | 1.46 (0.284) | 0.006 (0.003) | 0.018 (0.027) | 0.024 (0.029) |
| b_4c_90_0_unb_noise_2_4000 | 0.997 (0.001) | 0.879 (0.002) | 5.144 (0.455) | 1.231 (0.054) | 0.002 (0.002) | 0.005 (0.009) | 0.007 (0.01) |
| b_4c_90_4_bal_20_200 | 1 (0) | 0.875 (0.01) | 3.803 (0.439) | 0.766 (0.143) | 0.012 (0.012) | 0.003 (0.007) | 0.015 (0.015) |
| b_4c_90_4_bal_20_4000 | 1 (0) | 0.889 (0.002) | 2.683 (0.119) | 0.819 (0.046) | 0.007 (0.007) | 0.001 (0.003) | 0.008 (0.008) |
| b_4c_90_4_bal_2_200 | 0.985 (0.006) | 0.882 (0.008) | 2.591 (0.49) | 0.929 (0.168) | 0.008 (0.007) | 0.004 (0.009) | 0.011 (0.013) |
| b_4c_90_4_bal_2_4000 | 0.99 (0.001) | 0.894 (0.002) | 3.117 (0.217) | 0.905 (0.048) | 0.004 (0.005) | 0.001 (0.003) | 0.005 (0.006) |
| b_4c_90_4_bal_noise_20_200 | 1 (0) | 0.871 (0.01) | 3.33 (0.264) | 1.135 (0.185) | 0.006 (0.005) | 0.01 (0.018) | 0.016 (0.02) |
| b_4c_90_4_bal_noise_20_4000 | 1 (0) | 0.892 (0.002) | 7.11 (0.511) | 1.123 (0.048) | 0.002 (0.002) | 0.004 (0.008) | 0.006 (0.009) |
| b_4c_90_4_bal_noise_2_200 | 0.995 (0.003) | 0.872 (0.011) | 4.126 (1.047) | 1.211 (0.17) | 0.005 (0.004) | 0.011 (0.019) | 0.016 (0.021) |
| b_4c_90_4_bal_noise_2_4000 | 0.996 (0) | 0.892 (0.002) | 4.715 (0.353) | 1.141 (0.045) | 0.002 (0.002) | 0.004 (0.009) | 0.006 (0.009) |
| b_4c_90_4_unb_20_200 | 1 (0) | 0.864 (0.012) | 3.873 (0.398) | 0.732 (0.138) | 0.012 (0.012) | 0.002 (0.006) | 0.014 (0.014) |
| b_4c_90_4_unb_20_4000 | 1 (0) | 0.881 (0.002) | 2.764 (3.132) | 0.794 (0.043) | 0.007 (0.007) | 0 (0.002) | 0.007 (0.008) |
| b_4c_90_4_unb_2_200 | 0.984 (0.006) | 0.873 (0.01) | 2.41 (0.504) | 0.888 (0.156) | 0.008 (0.007) | 0.003 (0.007) | 0.011 (0.011) |
| b_4c_90_4_unb_2_4000 | 0.99 (0.001) | 0.886 (0.002) | 3.206 (0.24) | 0.88 (0.044) | 0.004 (0.004) | 0.001 (0.003) | 0.005 (0.005) |
| b_4c_90_4_unb_noise_20_200 | 1 (0) | 0.861 (0.013) | 3.355 (0.24) | 1.116 (0.178) | 0.006 (0.005) | 0.008 (0.016) | 0.014 (0.018) |
| b_4c_90_4_unb_noise_20_4000 | 1 (0) | 0.884 (0.002) | 7.237 (0.265) | 1.106 (0.049) | 0.002 (0.002) | 0.003 (0.007) | 0.005 (0.007) |
| b_4c_90_4_unb_noise_2_200 | 0.994 (0.003) | 0.863 (0.013) | 4.005 (1.206) | 1.182 (0.176) | 0.005 (0.004) | 0.009 (0.017) | 0.014 (0.019) |
| b_4c_90_4_unb_noise_2_4000 | 0.996 (0.001) | 0.885 (0.002) | 4.62 (0.387) | 1.123 (0.048) | 0.002 (0.001) | 0.003 (0.007) | 0.005 (0.008) |

## A2 Figures

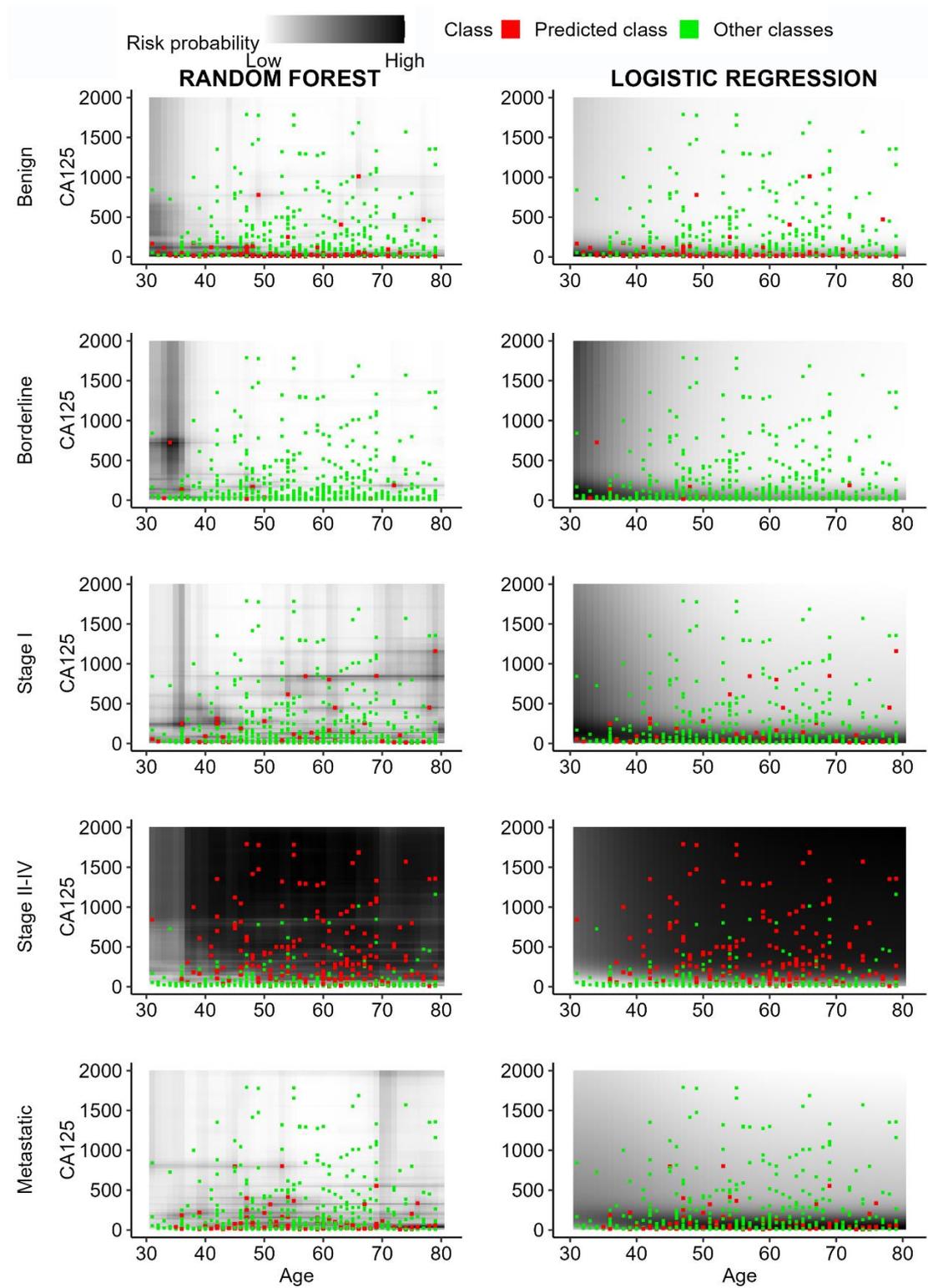

**Figure S1** Random forest and logistic regression probability estimation in data space for ovarian malignancy diagnosis with random forest (left) and multinomial logistic regression (right). Squares refer to train cases.

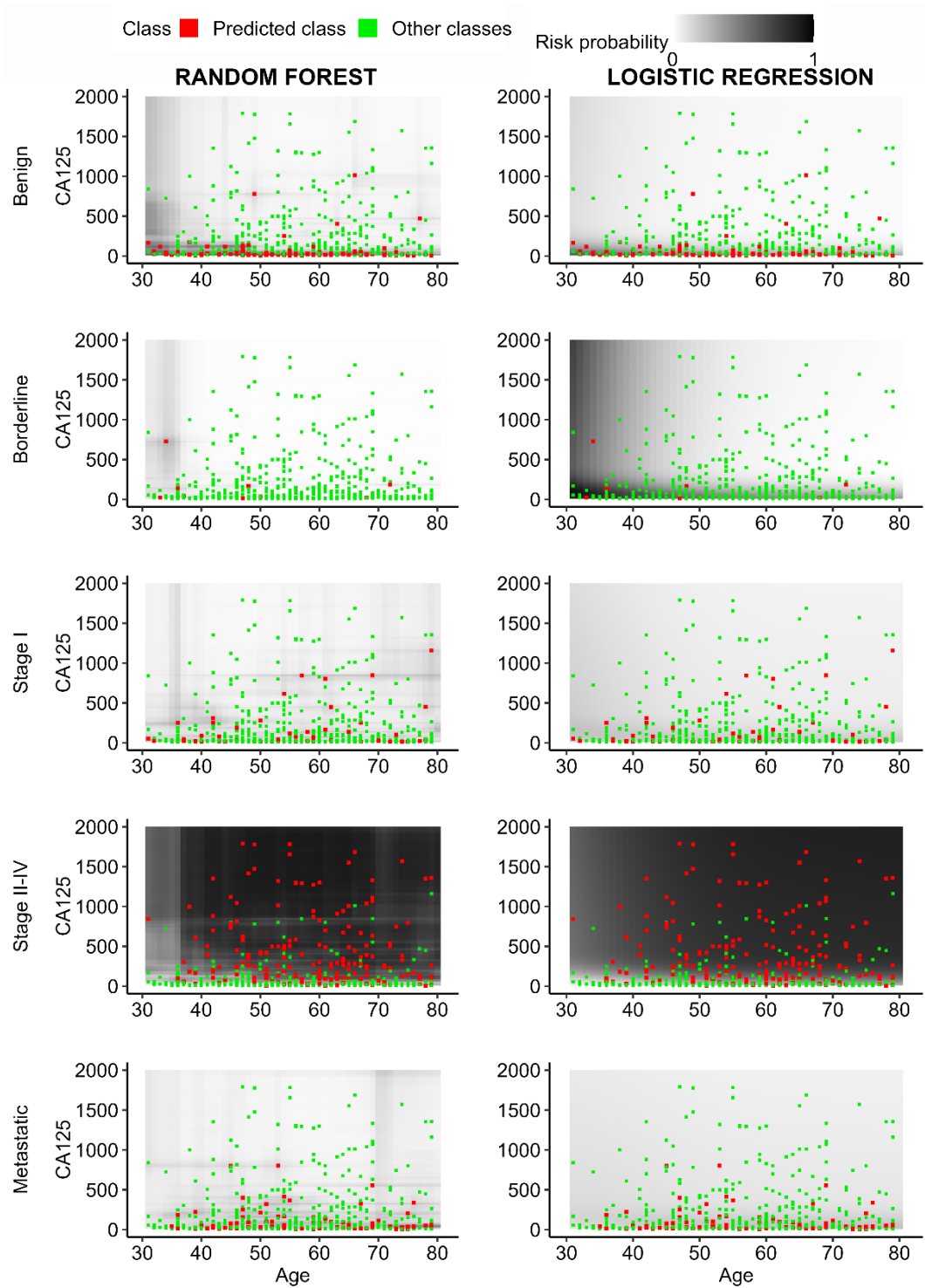

**Figure S2** Random forest and logistic regression probability estimation in data space for ovarian malignancy diagnosis with random forest (left) and multinomial logistic regression (right) using same scale for all panels. Squares refer to train cases.

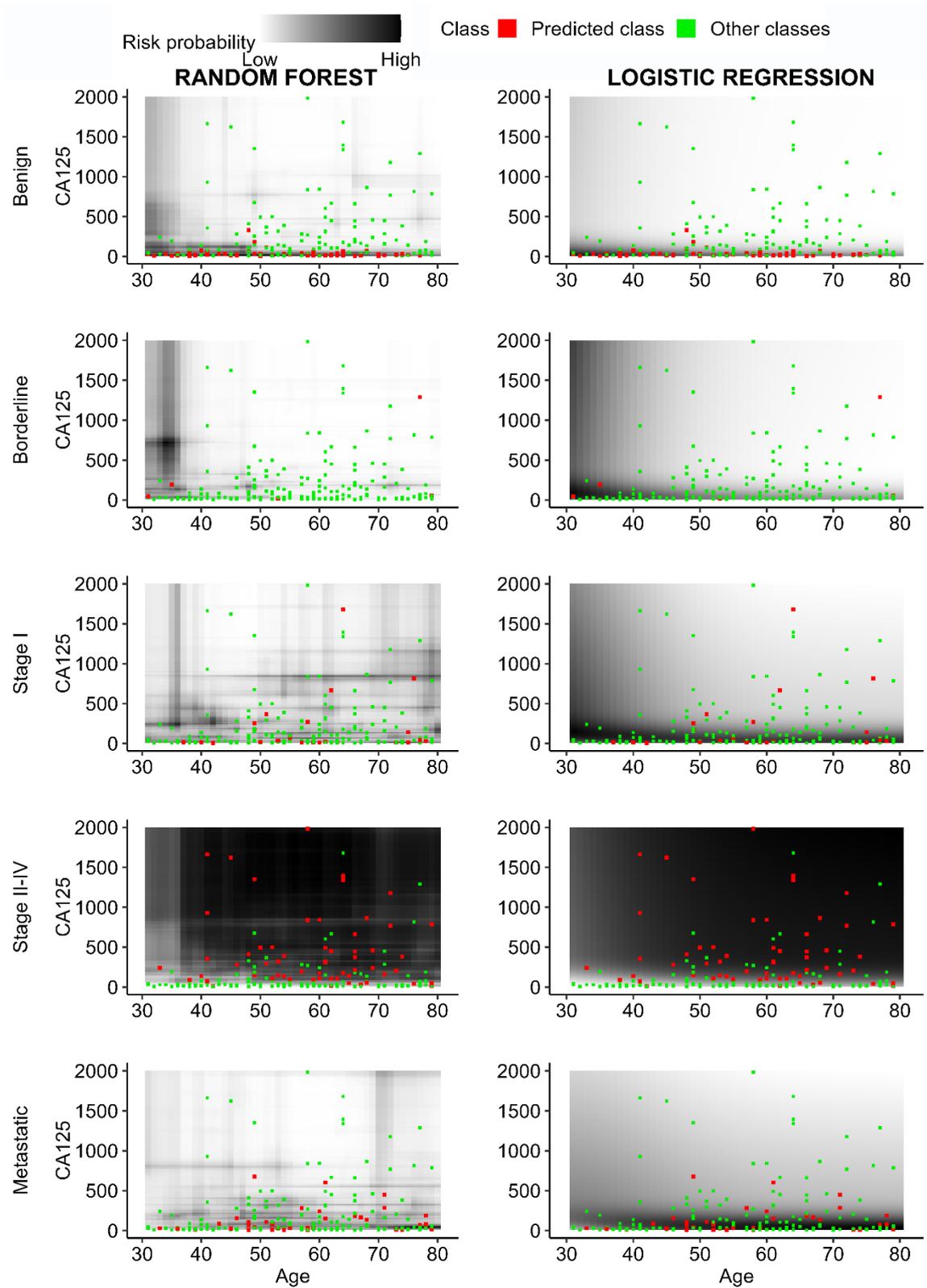

**Figure S3** Random forest and logistic regression probability estimation in data space for ovarian malignancy diagnosis with random forest (left) and multinomial logistic regression (right). Squares refer to test cases.

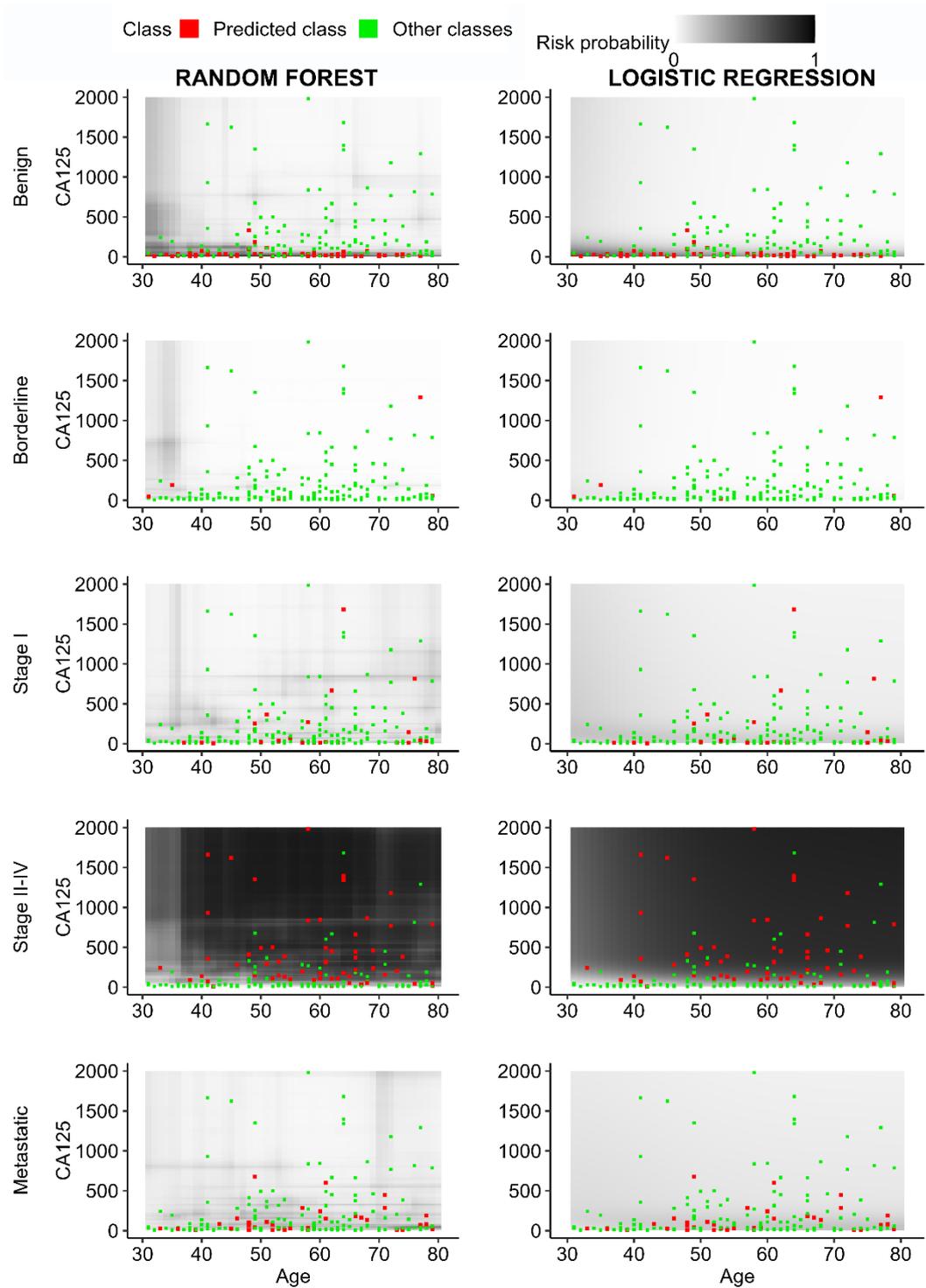

**Figure S4** Random forest and logistic regression probability estimation in data space for ovarian malignancy diagnosis with random forest (left) and multinomial logistic regression (right) using same scale for all panels. Squares refer to test cases.

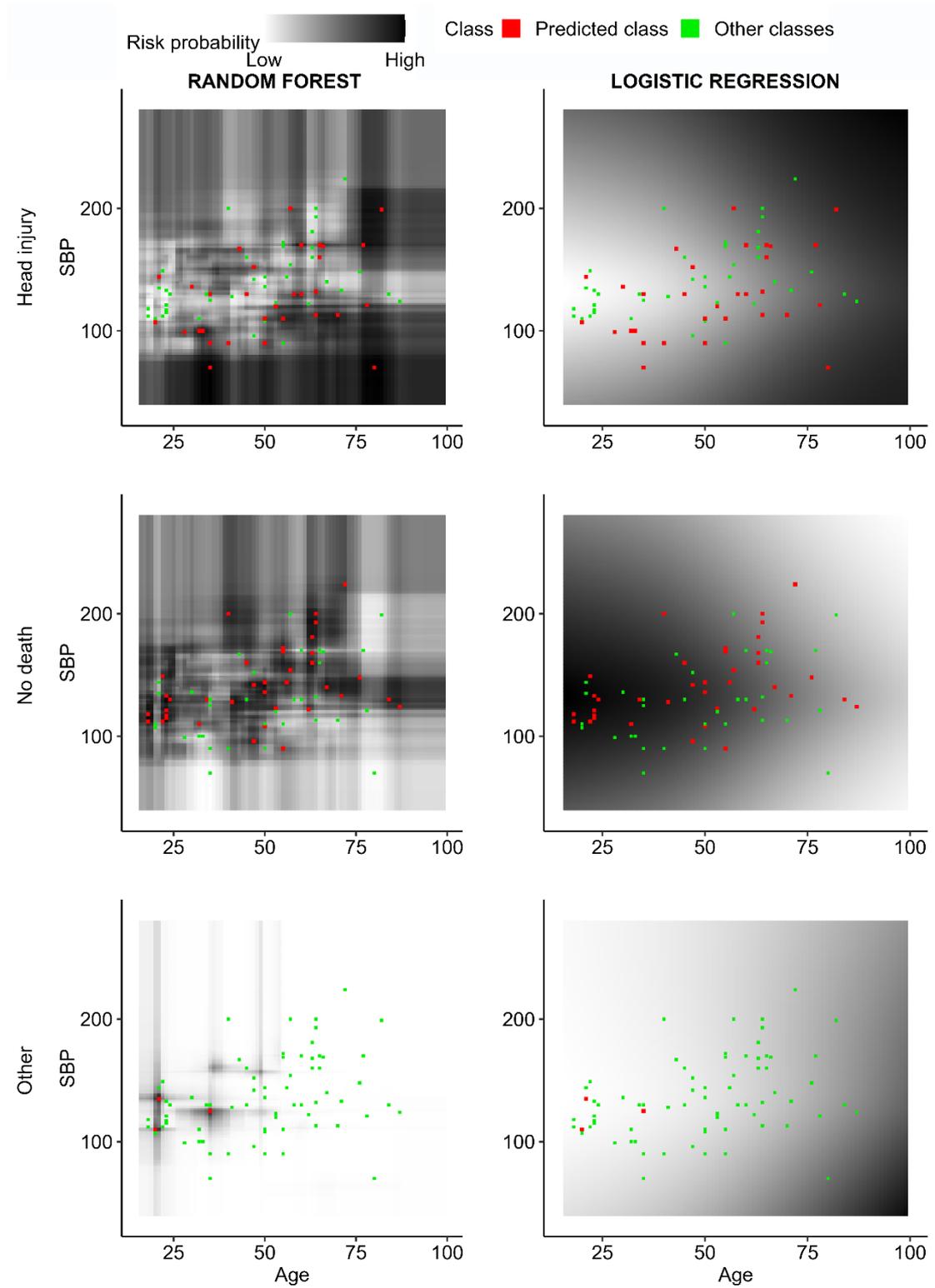

**Figure S5** Random forest and logistic regression probability estimation in data space for 3 possible outcomes in CRASH 3 dataset. Squares refer to train cases.

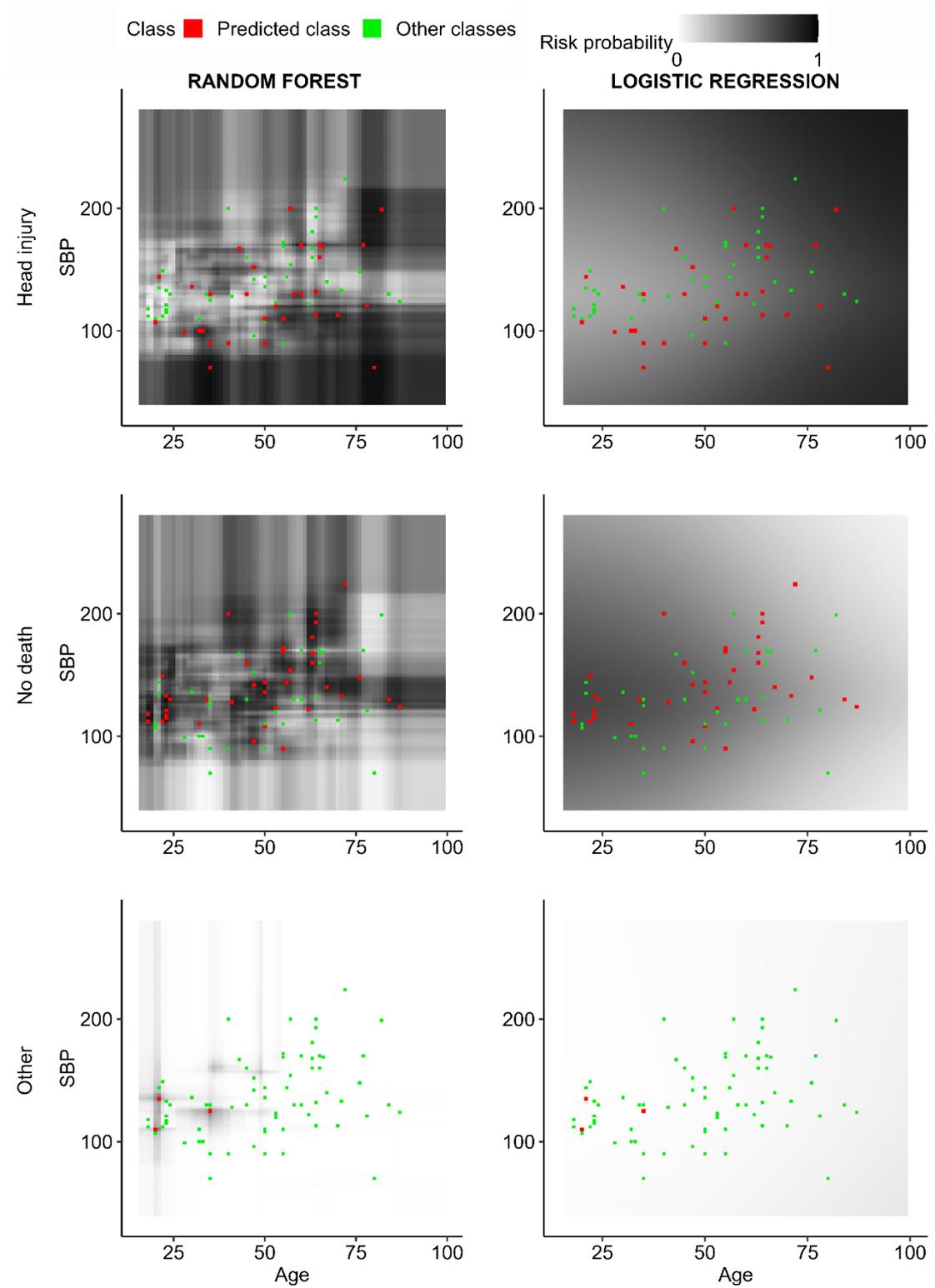

**Figure S6** Random forest and logistic regression probability estimation in data space for 3 possible outcomes in CRASH 3 dataset using same scale for all panels. Squares refer to train cases.

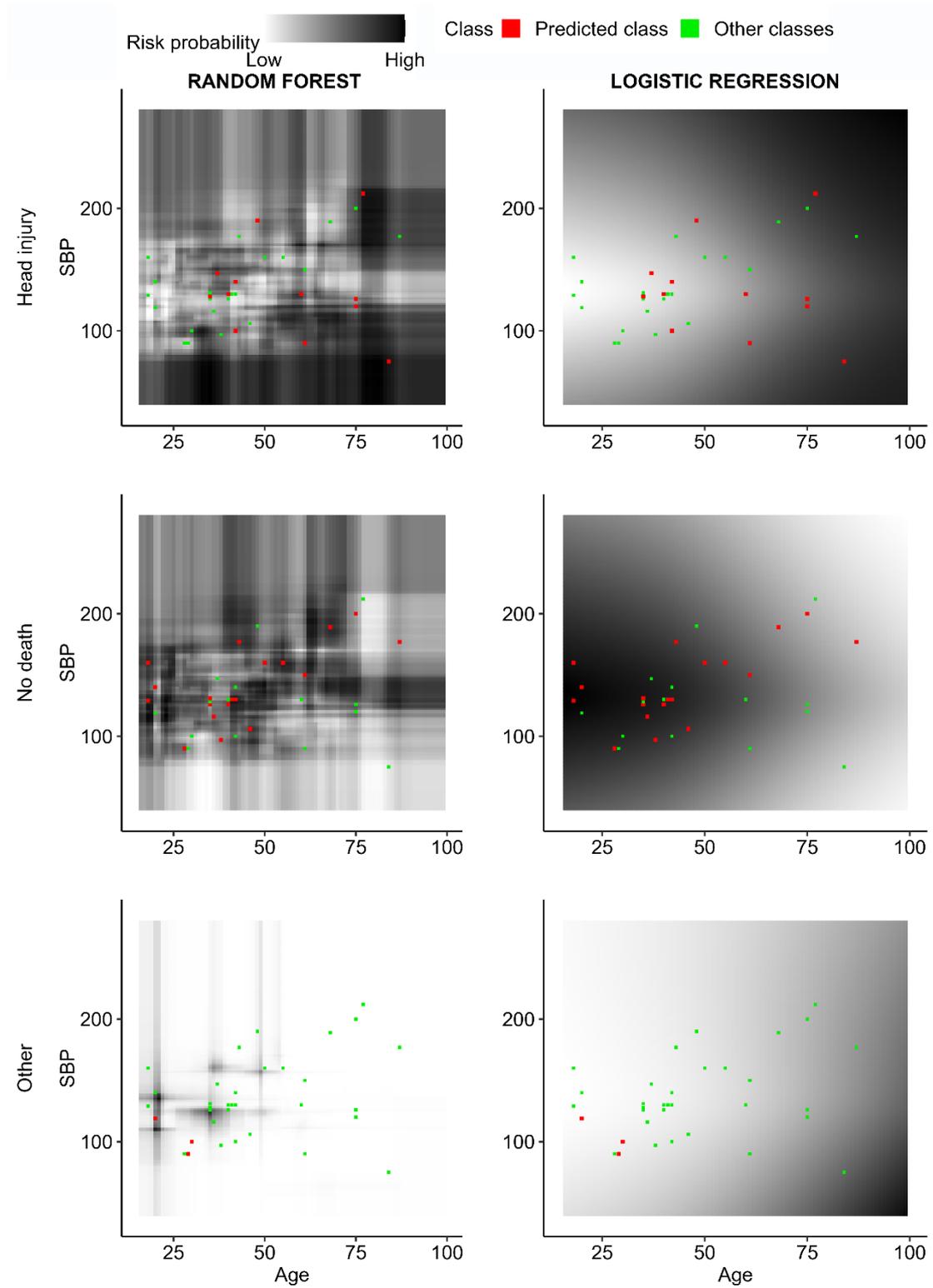

**Figure S7** Random forest and logistic regression probability estimation in data space for 3 possible outcomes in CRASH 3 dataset. Squares refer to test cases.

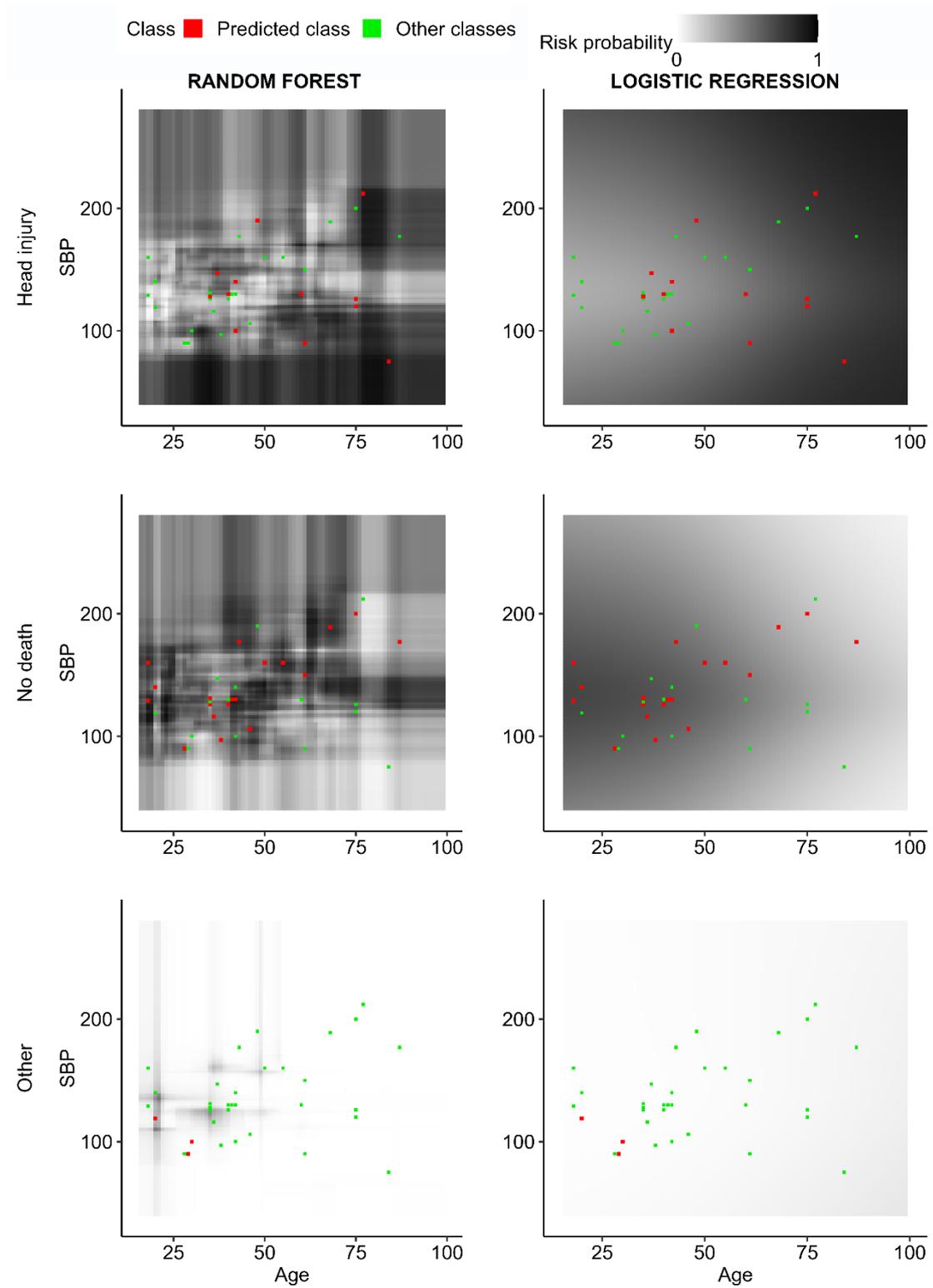

**Figure S8** Random forest and logistic regression probability estimation in data space for 3 possible outcomes in CRASH 3 dataset. Squares refer to test cases using same scale for all panels.

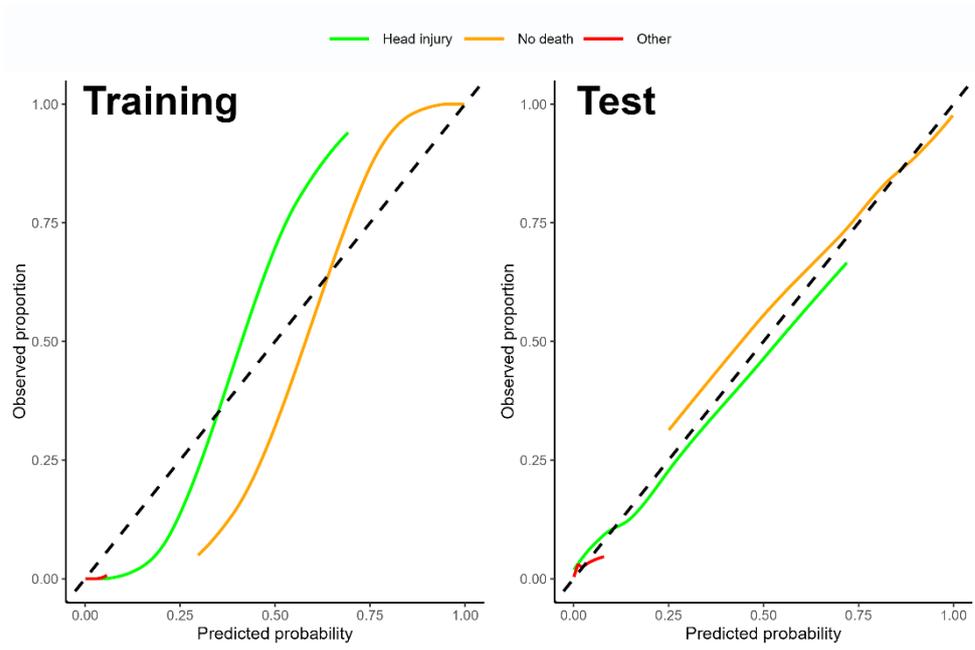

**Figure S9** Multinomial calibration plots of CRASH training and test data. Observed proration is estimated with a LOESS model. The plot shows only predicted probabilities between quantiles 5th and 95th.

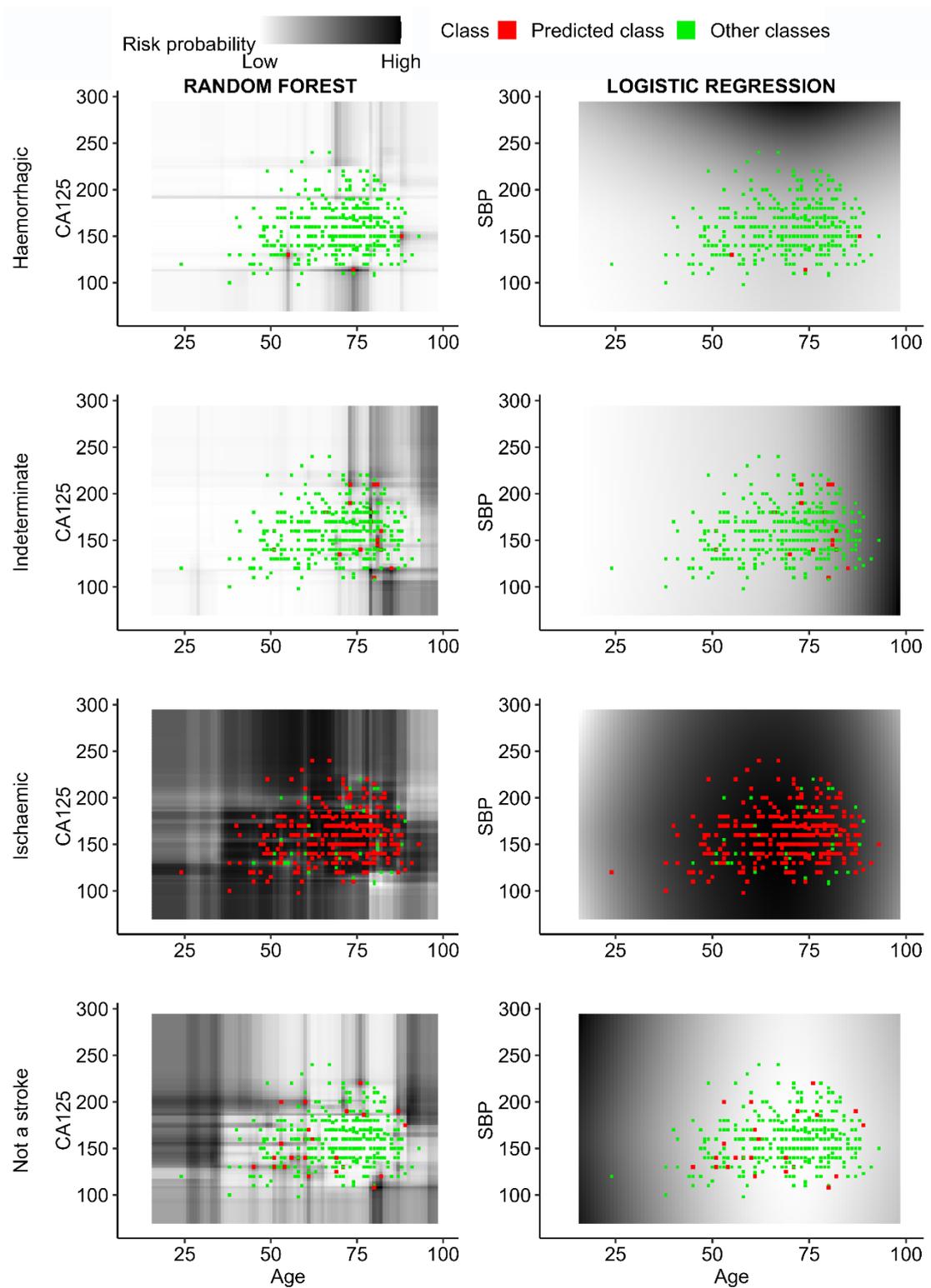

**Figure S10** Random forest and logistic regression probability estimation in data space for 4 possible type of strokes in IST dataset. Squares refer to training cases.

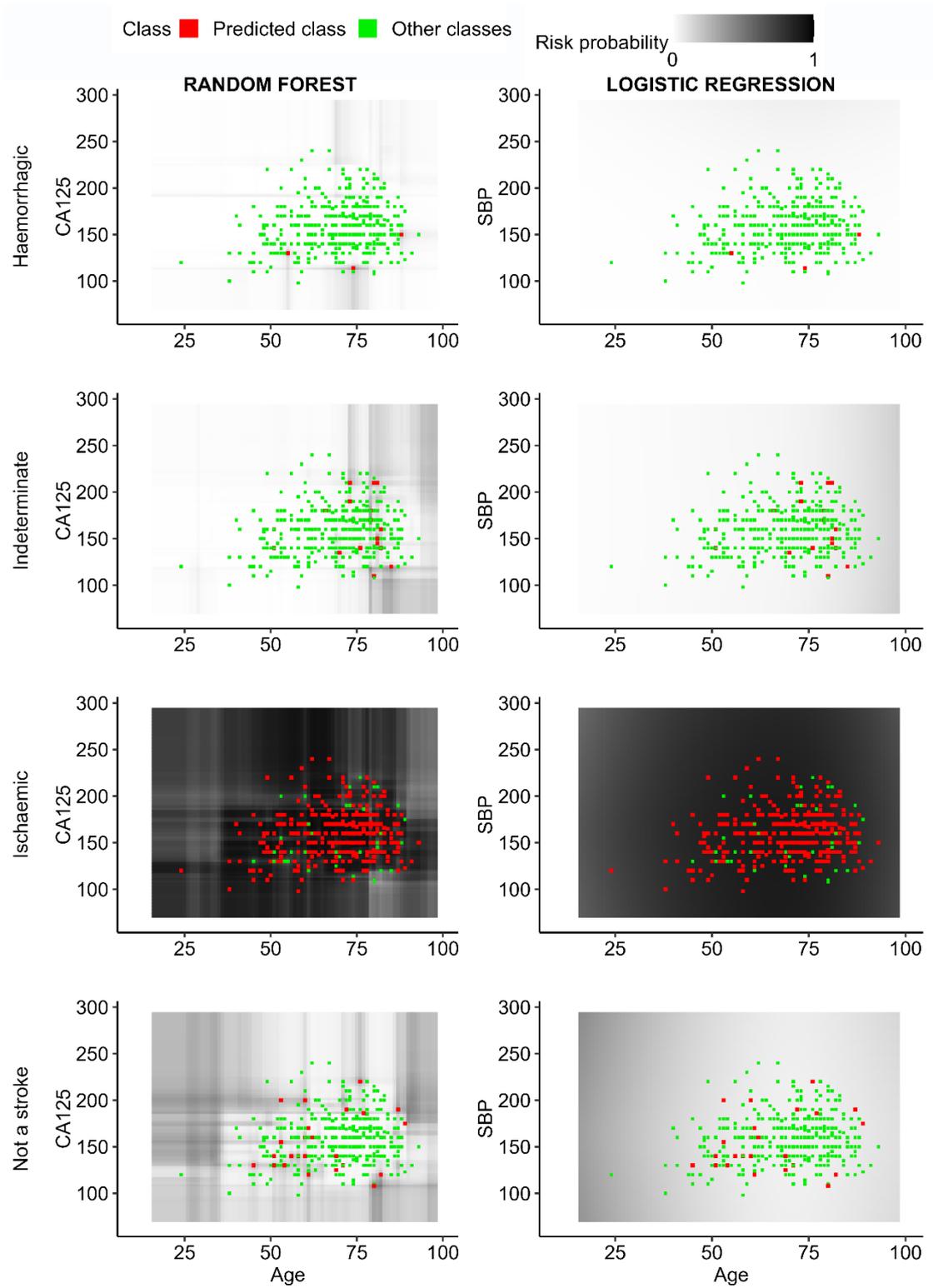

**Figure S11** Random forest and logistic regression probability estimation in data space for 4 possible type of strokes in IST dataset. Squares refer to training cases using same scale for all panels.

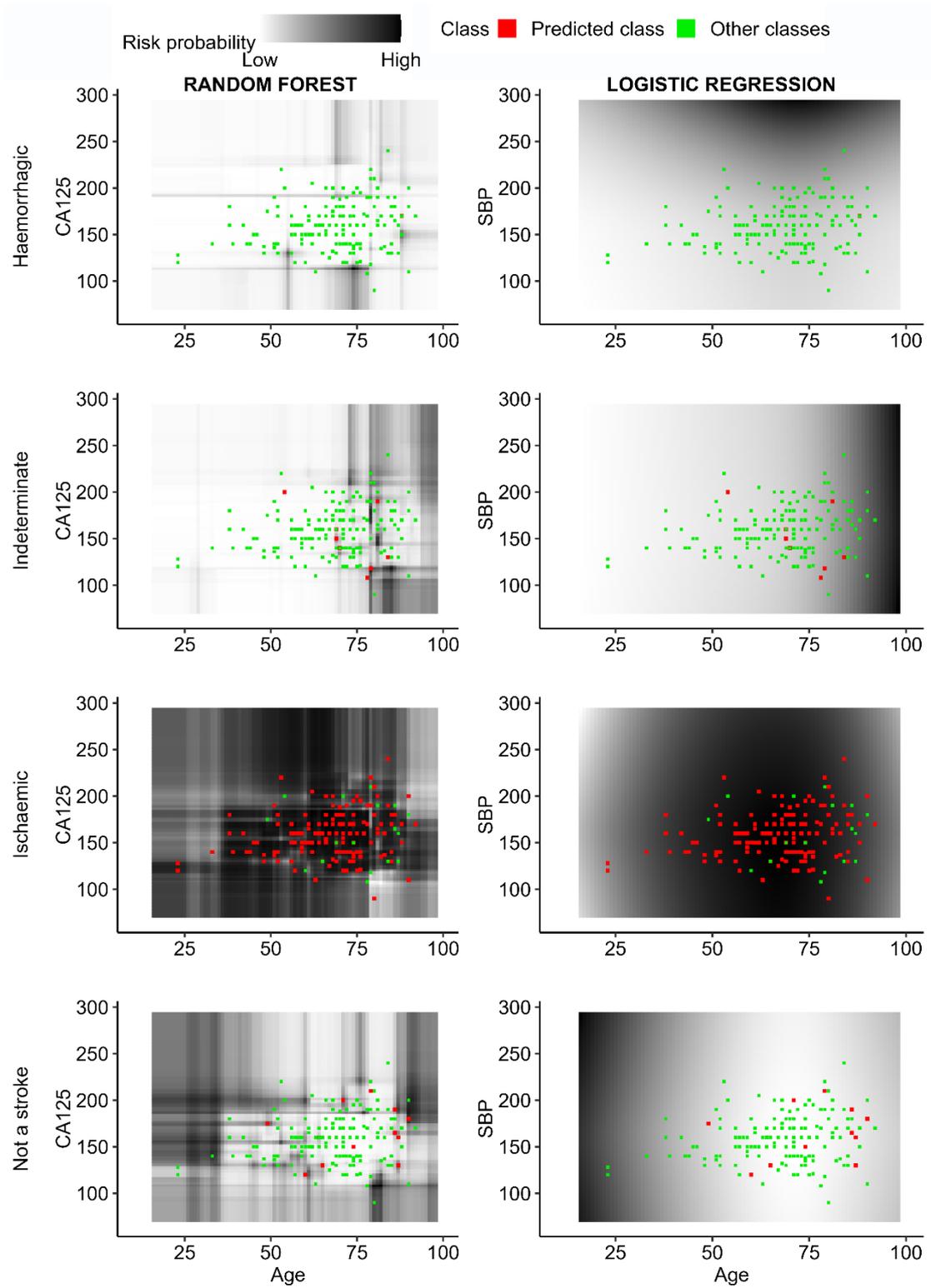

**Figure S12** Random forest and logistic regression probability estimation in data space for 4 possible type of strokes in IST dataset. Squares refer to test cases.

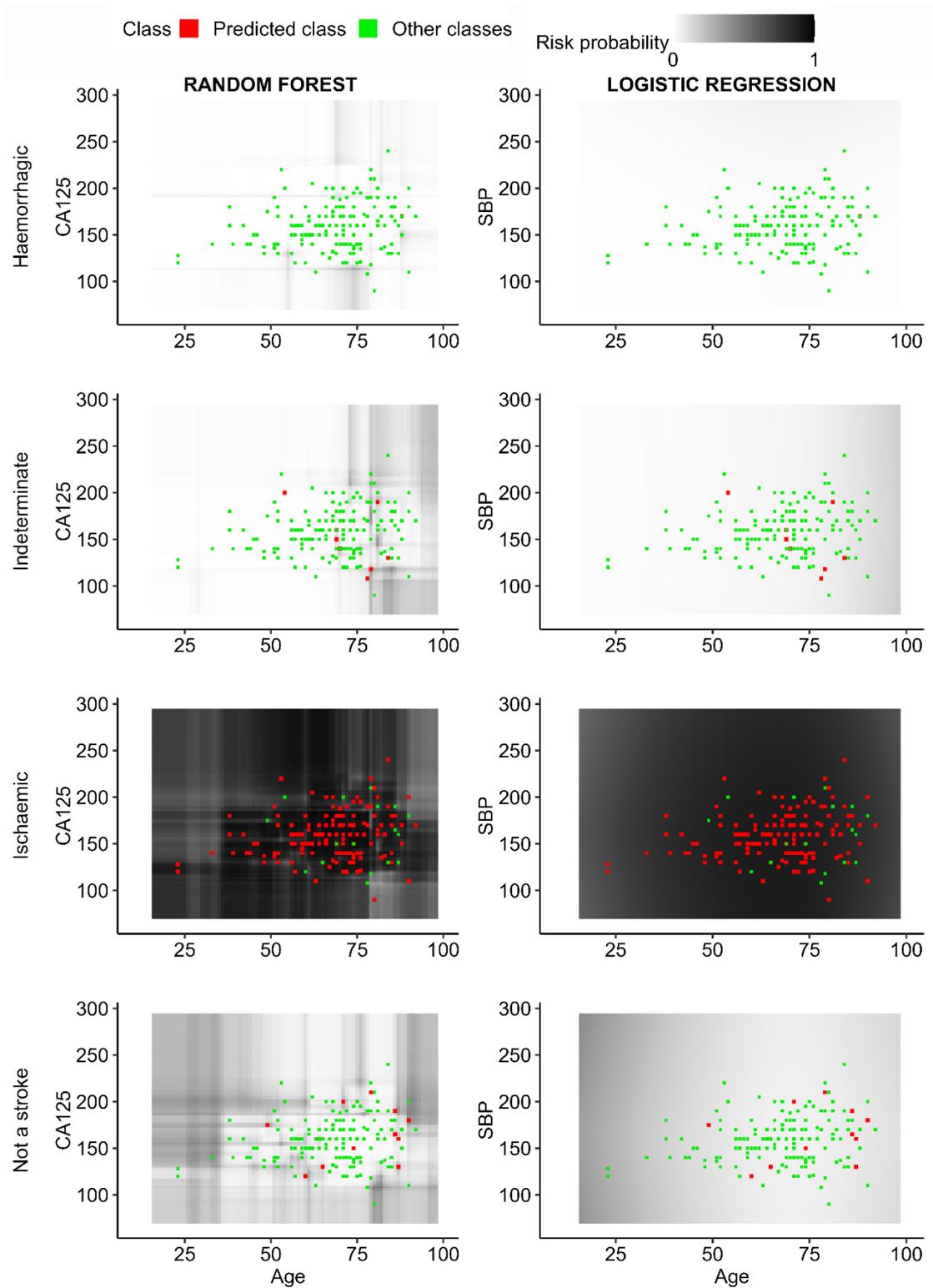

**Figure S13** Random forest and logistic regression probability estimation in data space for 4 possible type of strokes in IST dataset using same scale for all panels. Squares refer to test cases.

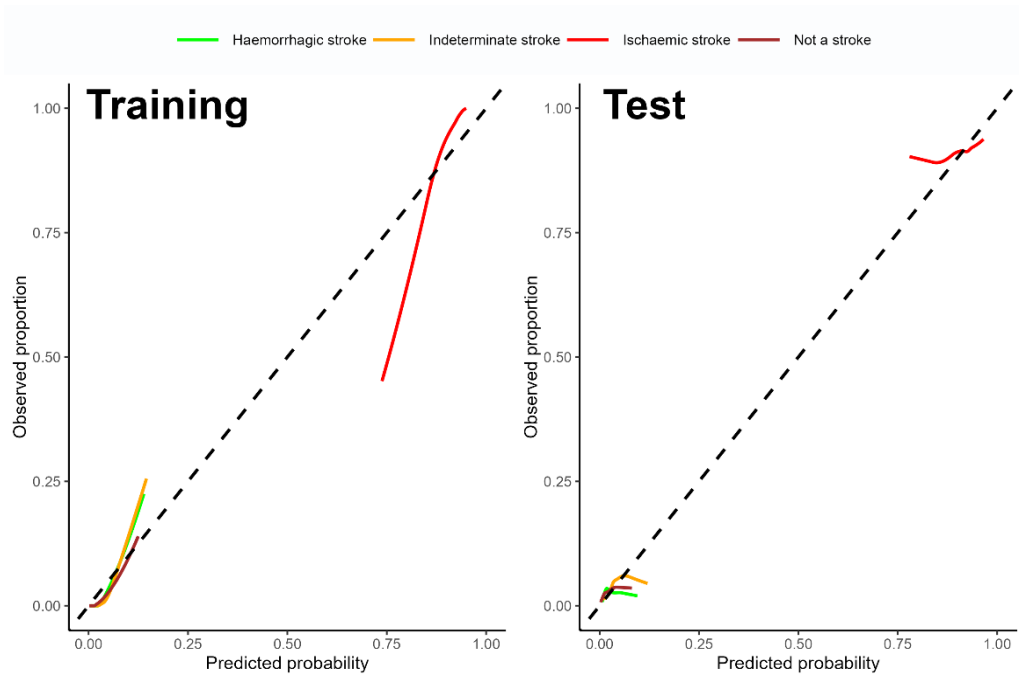

**Figure S14** Calibration plot in training for IST dataset. Observed proration is estimated with a LOESS model. The plot shows only predicted probabilities between quantiles 5th and 95th.

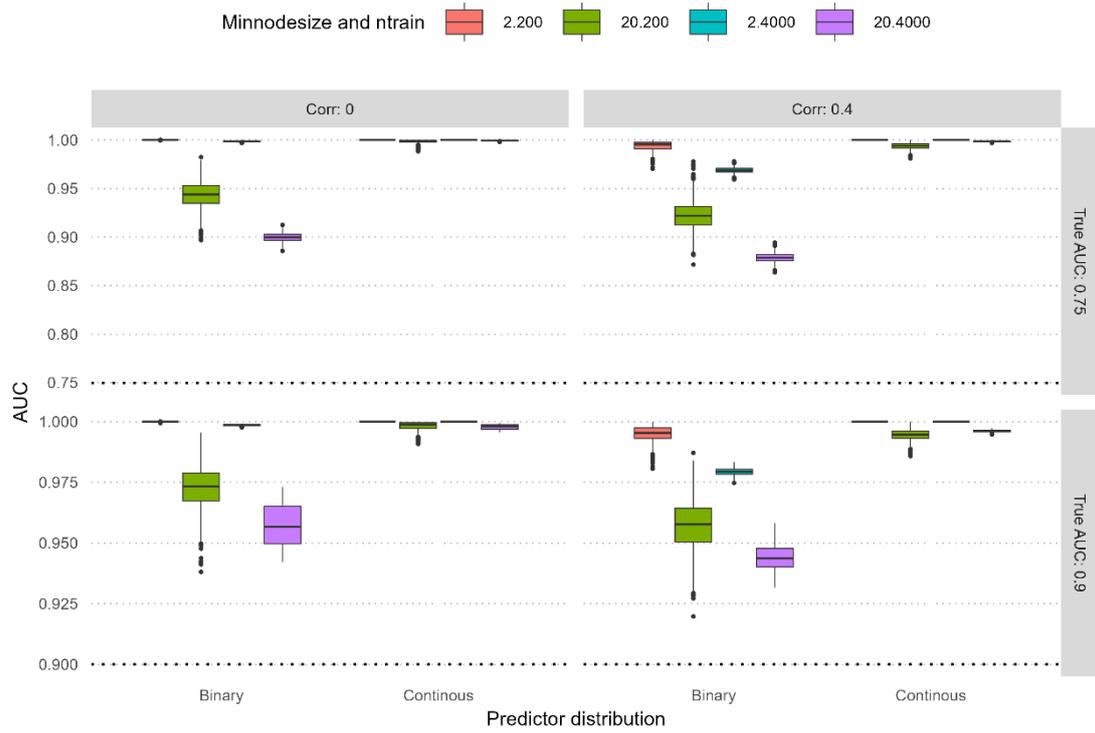

**Figure S15** Training AUC by simulation factors and modelling hyperparameters in scenarios with noise. Scenarios are aggregated by strength.

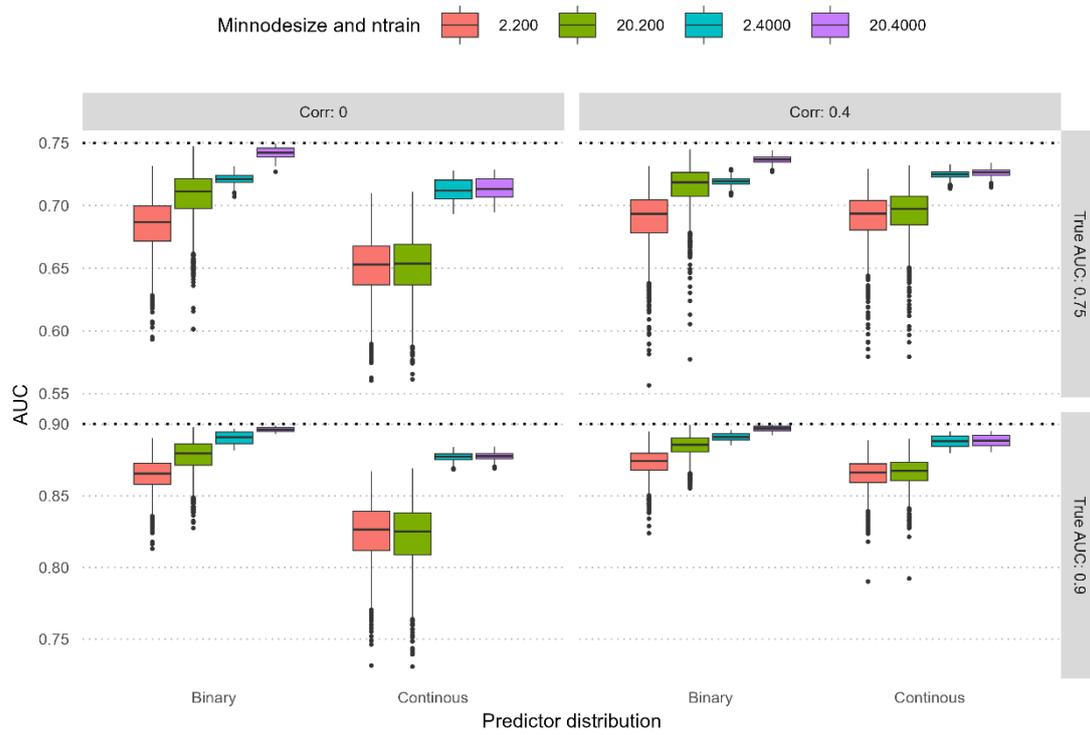

**Figure S16** Test AUC by simulation factors and modelling hyperparameters in scenarios with noise. Scenarios are aggregated by strength.

All

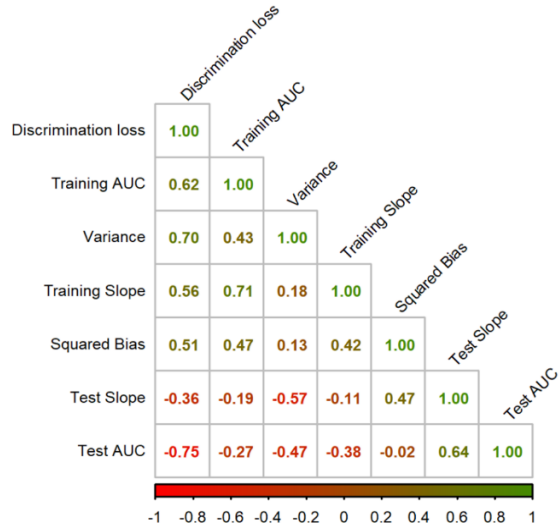

AUC = 0.9

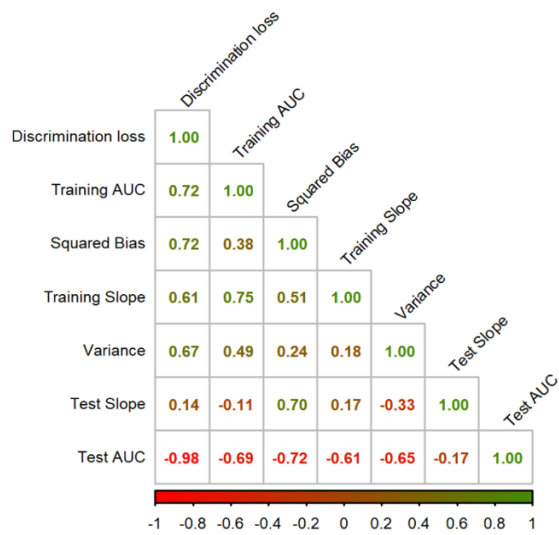

AUC = 0.75

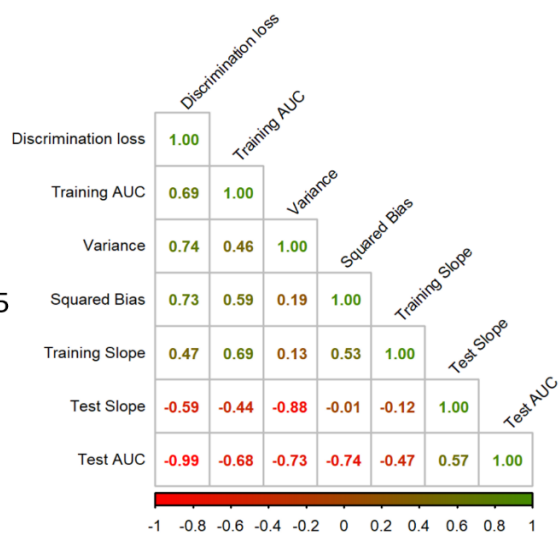

**Figure S17** Spearman correlations of principal metrics across all scenarios, scenarios with true AUCs 0.9 and scenarios with true AUC 0.75.

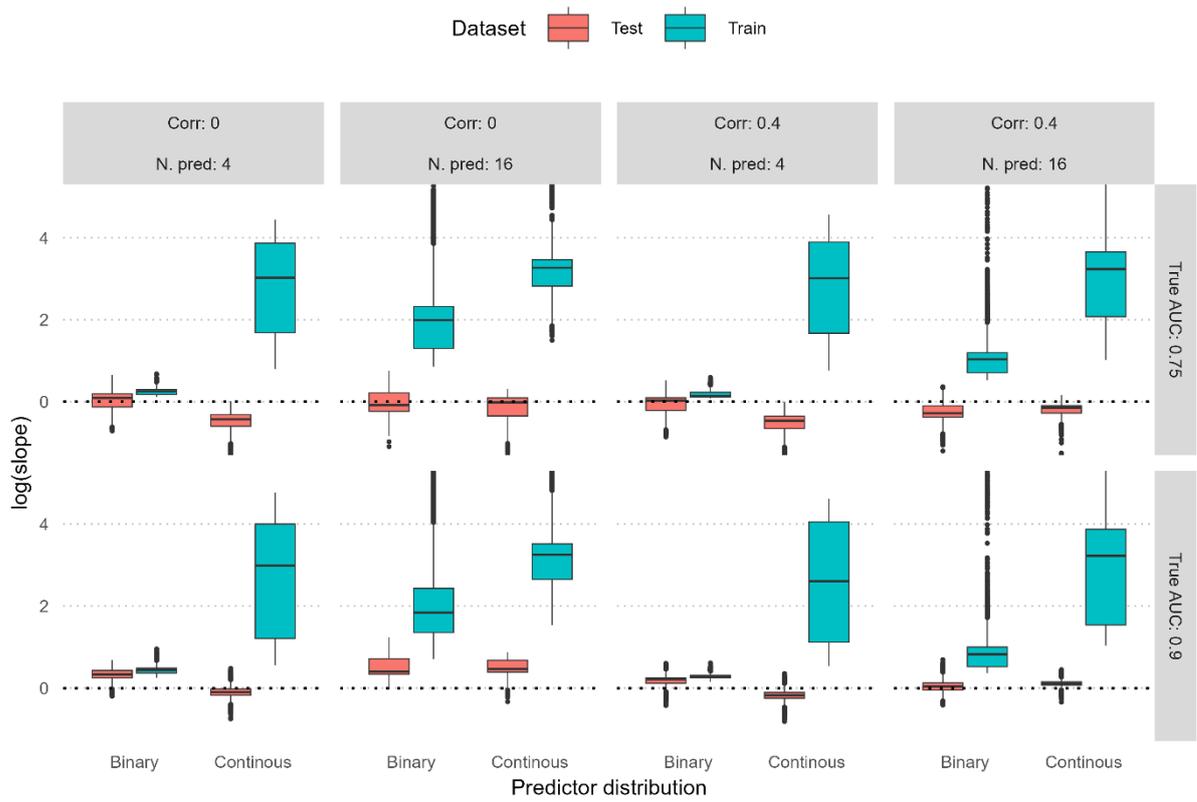

**Figure S18** Train and test calibration log(slope) in scenarios without noise. Scenarios are summarised by simulation factors that had minor effect.

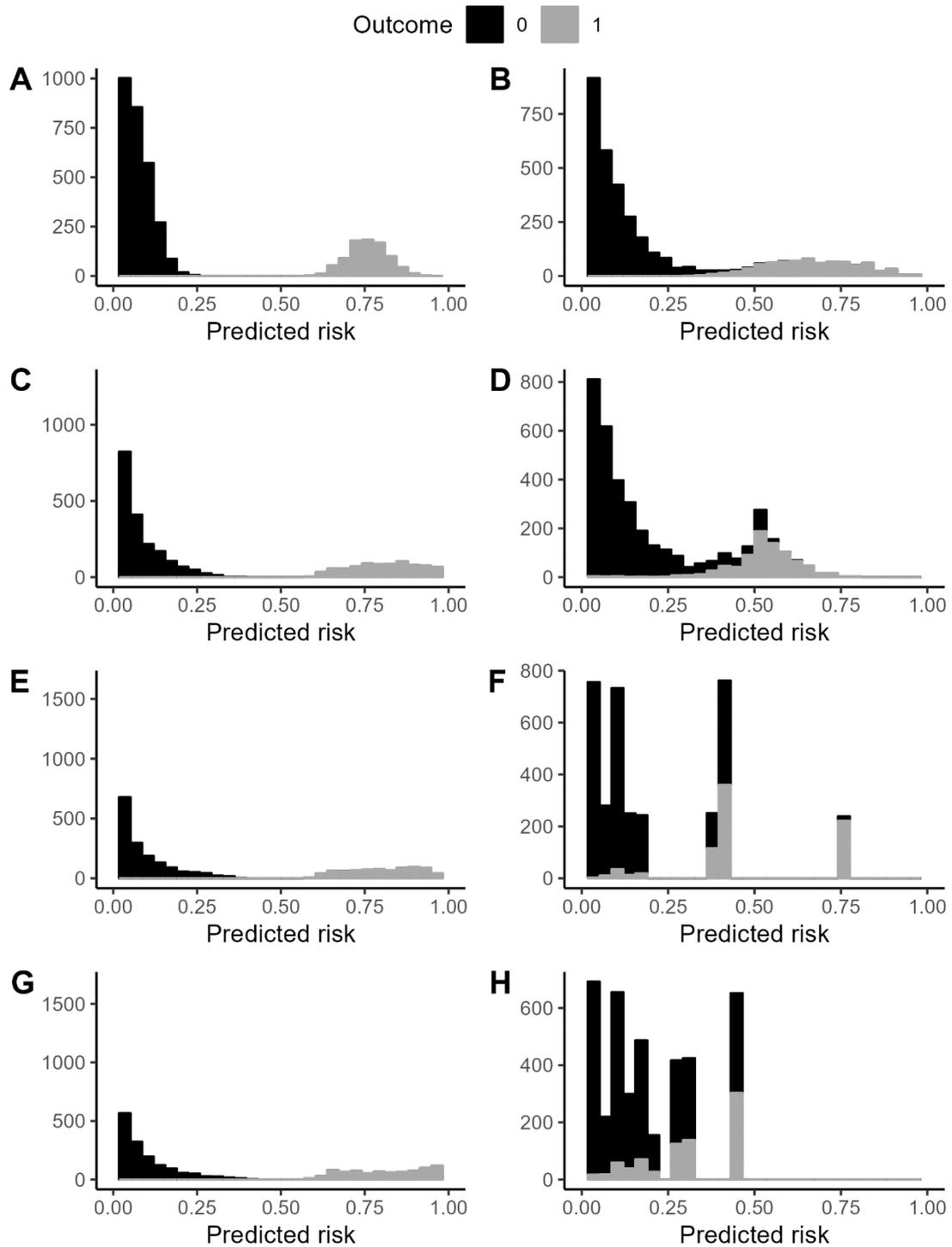

**Figure S19** Histogram of predicted probabilities in training for different simulation scenarios. Minimum node size was always 2 and training sample size 4000. Left column presents scenarios with contionus preductors and right column binary predictors. The scenarios were A (b_16c_90_0_bal), B (b_16b_90_0_bal), C(b_16c_90_4_bal), D(b_16b_75_4_bal), E (b_4c_90_0_bal), F (b_4b_90_0_bal), G(b_4c_90_4_bal), H(b_4b_75_4_bal), see table S5 for details. Code is available in OSF illustration_1.R.

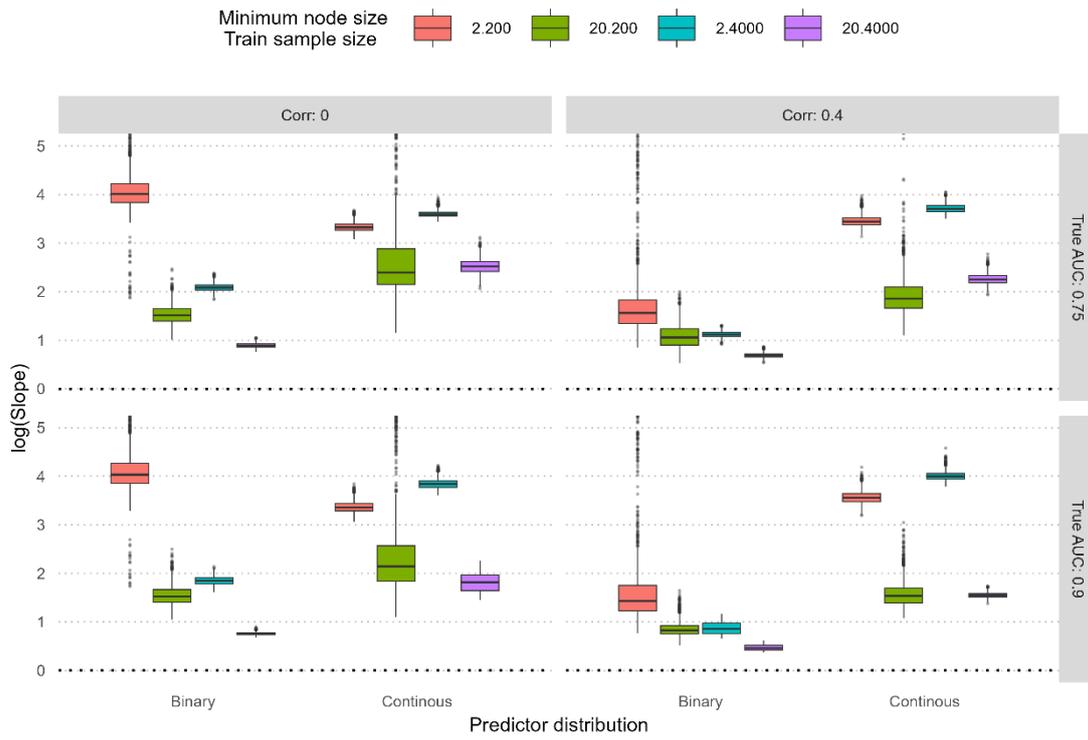

**Figure S20** Training set calibration log slope by simulation factors and modelling hyperparameters in scenarios with noise. Scenarios are aggregated by strength. Perfect calibration is 0.

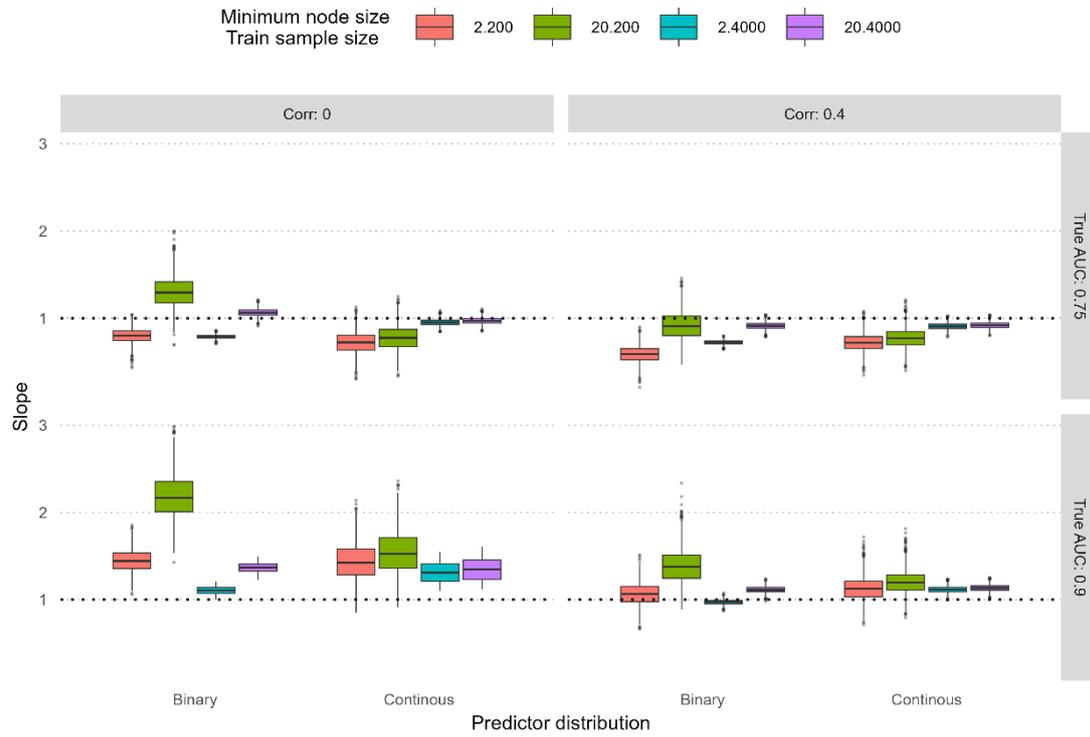

**Figure S21** Test set calibration slope by simulation factors and modelling hyperparameters in scenarios with noise. Scenarios are aggregated by strength. Perfect calibration is 1.

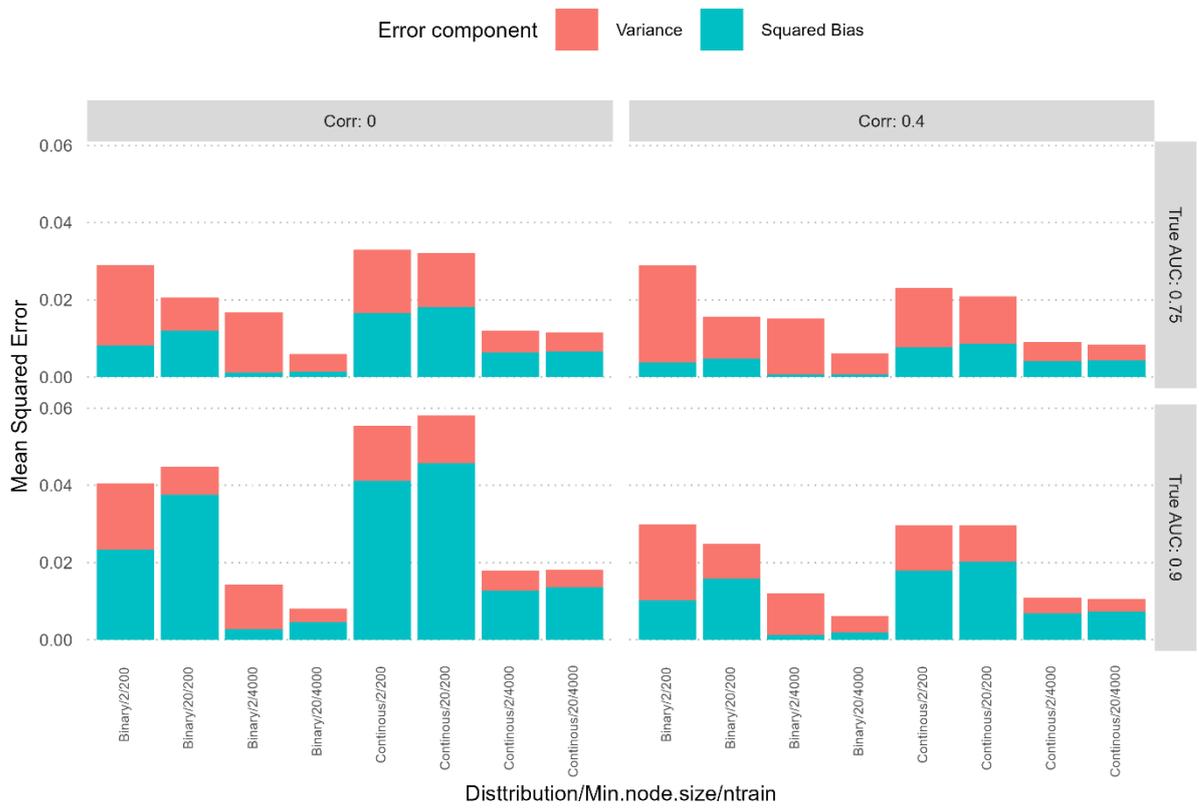

**Figure S22**   Mean squared error across scenarios with noise aggregated by strength.

### A.3 Simulation Algorithm

The code for the simulation function can be found in `OSF (https://osf.io/y5tqv/)` but here we will explain briefly how the data, probabilities and outcome were generated.

First we determine the regression coefficients so that all the simulation factors yield the desired AUC (see **Table S1**). In this work, this was done manually with trial error but we recommend to use automatised methods as the recently proposed iterative Bisection procedure (1) included in the *simstudy* `R package` or with the code provided in his paper.

1. **Input:**

    i. **num_samples**: Number of samples to generate (either 100.000 for test or 200 or 4000 for training).

    ii. **num_variables**: Number of predictor variables (16 or 4).

    iii. **coefficients**: An array of coefficients for the multinomial logistic regression model, including the intercept and predictor variable coefficients.

    iv. **correlation**: The desired correlation value between predictor variables (0 for no correlation).

    v. **binary_distribution**: Boolean flag indicating whether the outcome is binary.

2. **Generate Covariance Matrix:**

    i. If **correlation** is 0, use the identity matrix as the covariance matrix.

    ii. If **correlation** is not 0, create a matrix with all values as **correlation** except for the diagonal elements which are 1 (0.4).

3. **Generate Multivariate Normal Data:**

    i. Generate **num_samples** samples using a multivariate normal distribution with mean (μ) 0 and the covariance matrix obtained in step 2.

4. **Binarize Data (if binary distribution):**

    i. If **binary_distribution** is **true**, binarize the simulated data by setting values above a threshold (e.g., 0.5) to 1 and values below or equal to the threshold to 0.

5. **Calculate Linear Predictor:**

    i. For each sample, calculate the linear predictor by multiplying the values of the predictor variables by their corresponding coefficients and summing them, including the intercept.

6. **Calculate Predicted Probability:**

    i. For each linear predictor, calculate the predicted probability using the logistic function.

    $$p = \frac{1}{(1 + \exp(-LP)}$$

7. **Generate Binary Outcome:**

    i. Generate the outcome for each sample by drawing from a binomial distribution with n trials (1 trial) and probability of success (p) equal to the predicted probability.

8. **Output:**

    ii. **simulated_data**: The generated multivariate normal data.

    iii. **predicted_probability**: The array of predicted probabilities.

    iv. **binary_outcome**: The generated binary outcome based on the predicted probabilities.

**Note:** This algorithm outlines the step-by-step process for simulating data and calculating predicted probabilities for a binary outcome using multinomial logistic regression coefficients. The code that we provide also calculates the desired metrics.

## A.4 Simulation metrics

The predicted risks are denoted by $p_i$, true risks by $p_i^{true}$ and the true outcome by $y_i$ for patient i. Simulation are denoted by j.

Calibration slope: fitting a model $logit(Y) = intercept + p_i(slope)$

Calibration intercept: fitting a model $logit(Y) = intercept + offset(p_i)$

Perfectly separated $p_i$, where transformed to the closest possible value according to the number of trees and number of training samples as follows:

$$(p_i|p_i = 0) = \frac{0.5}{ntree \times ntrain} \text{ and } (p_i|p_i = 1) = \frac{ntree \times ntrain \times 0.5}{ntree \times ntrain}$$

Discrimination loss: $AUC_{true} - median(AUC_j)$

Bias: $mean(p_{ij}) - p_{ij}^{true}$

Variance: $mean\left(p_{ij} - mean(p_{ij})\right)^2$

Mean squared error (MSE): $bias^2 + variance$